\documentclass[aps,prl,twocolumn,superscriptaddress,a4paper,english,longbibliography]{revtex4-2}

\usepackage[table,xcdraw]{xcolor}
\usepackage{times}
\usepackage{color}
\usepackage[colorlinks=true,urlcolor=blue,citecolor=blue]{hyperref}
\usepackage{url}
\usepackage{soul}
\usepackage{graphicx}
\usepackage{amsmath}
\usepackage{subfigure}
\usepackage{xcolor}
\usepackage{amssymb}
\usepackage{latexsym}
\usepackage{array}
\usepackage{multirow}
\usepackage{hyperref}
\usepackage{float}
\usepackage{ulem}
\usepackage{appendix}
\usepackage{etoolbox}
\usepackage{pifont}
\usepackage{mhchem}
\usepackage{makecell}
\usepackage{placeins}

\usepackage{xcolor}
\usepackage[urlcolor=blue]{hyperref}      
\hypersetup{
    colorlinks = true,                    
    citecolor = {blue},
    linkcolor = {purple},
           }
\begin{document}	
	
\title{Universal scaling laws in quantum-probabilistic machine learning by tensor network towards interpreting representation and generalization powers}

\author{Sheng-Chen Bai}
\affiliation{Department of Physics, Capital Normal University, Beijing 100048, China}

\author{Shi-Ju Ran}
\email[Corresponding author: Shi-Ju Ran ]{sjran@cnu.edu.cn}
\affiliation{Department of Physics, Capital Normal University, Beijing 100048, China}

\date{\today}

\begin{abstract}
	Interpreting the representation and generalization powers has been a long-standing issue in the field of machine learning (ML) and artificial intelligence. This work contributes to uncovering the emergence of universal scaling laws in quantum-probabilistic ML. We take the generative tensor network (GTN) in the form of a matrix product state as an example and show that with an untrained GTN (such as a random TN state), the negative logarithmic likelihood (NLL) $L$ generally increases linearly with the number of features $M$, i.e., $L \simeq k M + const$. This is a consequence of the so-called ``catastrophe of orthogonality,'' which states that quantum many-body states tend to become exponentially orthogonal to each other as $M$ increases. We reveal that while gaining information through training, the linear scaling law is suppressed by a negative quadratic correction, leading to $L \simeq \beta M - \alpha M^2 + const$. The scaling coefficients exhibit logarithmic relationships with the number of training samples and the number of quantum channels $\chi$. The emergence of the quadratic correction term in NLL for the testing (training) set can be regarded as evidence of the generalization (representation) power of GTN. Over-parameterization can be identified by the deviation in the values of $\alpha$ between training and testing sets while increasing $\chi$. We further investigate how orthogonality in the quantum feature map relates to the satisfaction of quantum probabilistic interpretation, as well as to the representation and generalization powers of GTN. The unveiling of universal scaling laws in quantum-probabilistic ML would be a valuable step toward establishing a white-box ML scheme interpreted within the quantum probabilistic framework.
\end{abstract}
\maketitle

\textit{Introduction.---} Developing ``white-box'' (or interpretable) schemes of machine learning (ML) and artificial intelligence is a long-concerned issue that is key to ensuring robustness, stability, fairness, privacy protection, etc.~\cite{gilpin2018explaining,zhang2018visual,carvalho2019machine}. Among others, characterizing and interpreting representation and generalization powers, in other words, the abilities of ML models to handle both learned and unlearned samples, is a fundamental topic in both academic research and real-life applications.

Tensor network (TN) has been considered as a promising candidate for developing ``white-box'' ML due to its solid and systematic theoretical foundation in quantum information sciences and quantum mechanics (see, e.g., Refs.~\cite{SS16TNML, liu2019machine,han2018unsupervised,cheng2019tree, sun2020generative,Bai:100701,PhysRevE.107.L012103,MZZGR23ResMPS,10517663,PozasKerstjens2024privacypreserving} and a recent review in Ref.~\cite{doi:10.34133/icomputing.0061}). Originating from quantum many-body simulation, TN has been regarded as an interpretable and efficient numerical tool~\cite{doi:10.1080/14789940801912366,Cirac_2009,ORUS2014117,O19TNrev,RTPC+17TNrev,cirac2021matrix}, with its representation power characterized by the scaling laws of entanglement entropy (see, e.g.,~\cite{PhysRevLett.90.227902,PhysRevLett.102.255701,SWVC08MPSent,TOIL08EntScaling}). The representation power in quantum many-body simulation can be understood as how the simulation accuracy of the physical properties scales with the parameter complexity. For instance, matrix product state (MPS)~\cite{PVWC07MPSRev,cirac2021matrix} provides a faithful representation of the states satisfying the one-dimensional (1D) area law of entanglement entropy, offering a solid foundation for the efficient simulation of 1D quantum systems with local interactions by MPS.

In the field of ML, the representation and generalization powers can be characterized by the loss functions or prediction accuracies on the training and testing sets, respectively. Recent researches have uncovered the power laws in the scaling of the loss function in recursive neural networks~\cite{hestness2017deeplearningscalingpredictable} and Transformer models~\cite{kaplan2020scalinglawsneurallanguage}, which have been applied in different scenarios such as multi-modality~\cite{henighan2020scalinglawsautoregressivegenerative,Zhai_2022_CVPR}, compute-constrained environments~\cite{hoffmann2022trainingcomputeoptimallargelanguage}, data engineering~\cite{NEURIPS2022_7b75da9b,NEURIPS2023_9d89448b}, and reinforcement learning~\cite{pmlr-v202-gao23h}. There are also some pioneering works on revealing the scaling behaviors of TN-based ML~\cite{PhysRevE.107.L012103,Convy_2022}, which have preliminarily demonstrated the significance of this topic.

\begin{figure}[tbp]
	\centering
	\includegraphics[angle=0,width=0.72\linewidth]{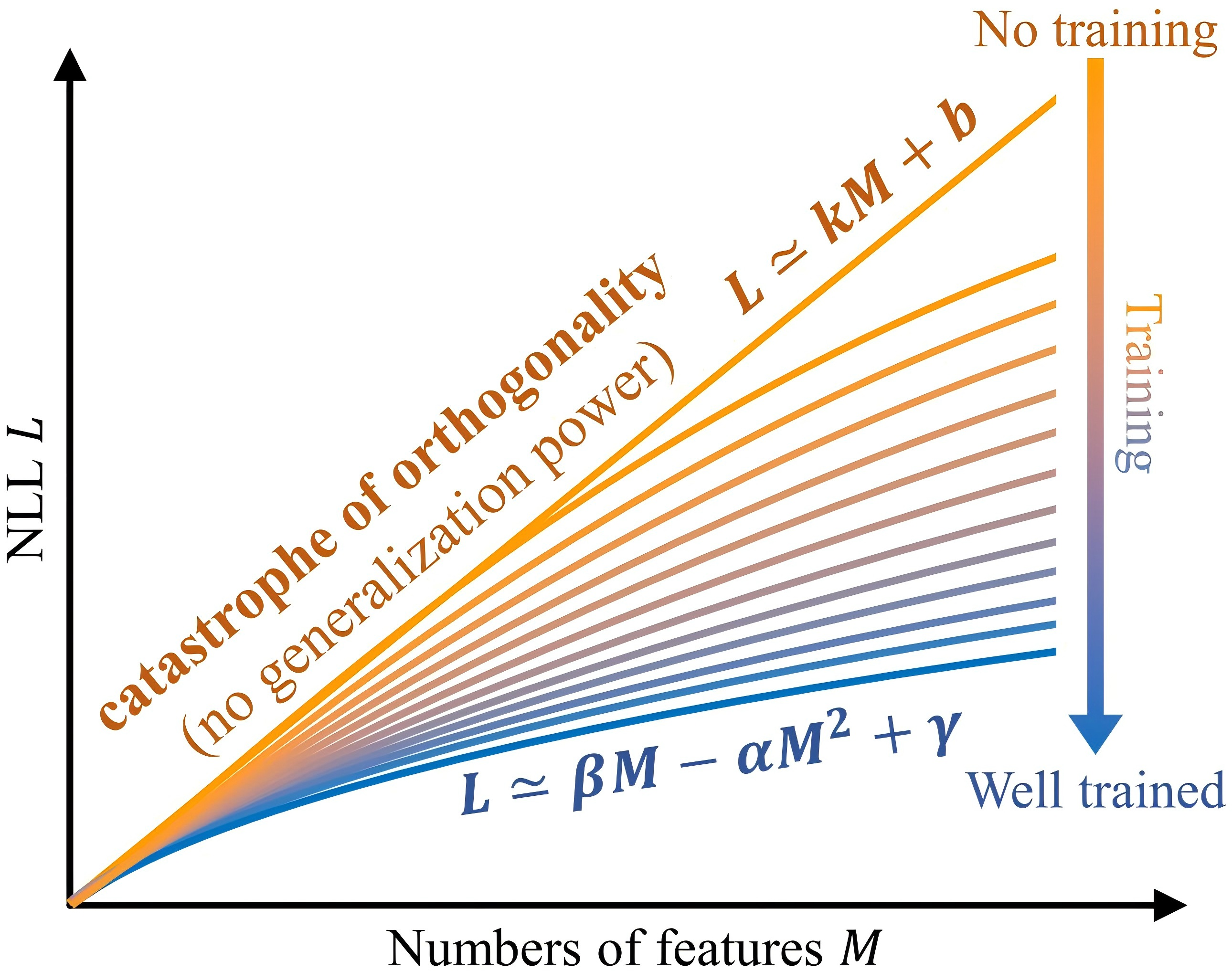}
	\caption{(Color online) The illustration of how the gain of information alters the scaling law of NLL in quantum-probabilistic ML. By training the GTN, the linear scaling law with respect to the feature number $M$ [Eq.~(\ref{eq-linear})], which is the result of the ``orthogonal catastrophe'' of quantum many-body states, is suppressed by the addition of a negative quadratic correction term [Eq.~(\ref{eq-corrected})].}
	\label{fig-idea}
\end{figure}

This work contributes to uncovering the scaling laws in quantum-probabilistic ML, which provides interpretations of the representation and generalization powers. The idea of quantum-probabilistic ML is to model the joint probabilistic distributions by quantum many-body states that satisfy Born's quantum probabilistic interpretation~\cite{doi:10.34133/icomputing.0061}. We take the generative TN (GTN) in the MPS form as an example~\cite{han2018unsupervised}. With a GTN that is irrelevant to the data (e.g., untrained or random), the negative logarithmic likelihood (NLL) $L$ universally obeys a linear scaling law with respect to the number of features $M$ as
\begin{eqnarray}
	L \simeq kM + b.
	\label{eq-linear}
\end{eqnarray}
This suggests the exponentially-vanishing probabilities of the samples obtained from the GTN, which is a consequence of the ``catastrophe of orthogonality (COO)'' for the quantum many-body states. By training the GTN, the acquired information suppresses the linear increase of NLL with a negative quadratic correction term as
\begin{eqnarray}
	L \simeq \beta M - \alpha M^2 + \gamma.
	\label{eq-corrected}
\end{eqnarray}
See Fig.~\ref{fig-idea} for an illustration.

The scaling coefficients in $L$ exhibit logarithmic scaling behaviors against the virtual dimension $\chi$ of the GTN. These coefficients' scaling laws are consistent with the logarithmic scaling of NLL versus $\chi$, which explains how the powers of GTN improve as the number of quantum channels increases. The deviation of the coefficients' scaling behaviors between the training and testing sets indicates over-parameterization. We further investigate how the orthogonality of the quantum feature map relates to the satisfaction of quantum probabilistic interpretation, as well as to the scaling laws, and to the representation and generalization powers of GTN.

\textit{Generative tensor network for quantum-probabilistic machine learning.---} The central idea of quantum-probabilistic machine learning is to represent the joint probabilistic distribution of $M$ features (denoted as $P(\boldsymbol{x})$ with $\boldsymbol{x} = (x_1, x_2, \dots, x_M)$) by a quantum many-body state (denoted as $|\Psi\rangle$) consisting of $M$ spins (or qubits). The state is related to $P(\boldsymbol{x})$ via protective measurement as
\begin{eqnarray}
	P(\boldsymbol{x}) = \left| \langle \boldsymbol{x}|\Psi\rangle \right|^2,
	\label{eq-P}
\end{eqnarray}
where the product state $| \boldsymbol{x}\rangle \equiv |x_1, x_2, ..., x_M\rangle$ is obtained from the sample $\boldsymbol{x}$ by adopting the so-called quantum feature map (QFM).

The definition of QFM can be flexible. Here, we consider a widely-used one that reads
\begin{equation}
	|x_1, x_2, ..., x_M\rangle = \prod_{m=1}^M \left[ \cos \frac{\theta \pi x_m}{2} |0\rangle + \sin \frac{\theta \pi x_m}{2} |1\rangle \right],
	\label{eq-QFM}
\end{equation}
with $\theta$ a hyper-parameter that tunes the orthogonality, and $|0\rangle$ and $|1\rangle$ being the eigenstates of the Pauli matrix $\boldsymbol{\sigma}^z$~\cite{SS16TNML}. The above formulation means that with a well-trained $|\Psi\rangle$, the probability of any given sample $\boldsymbol{x}$ can be obtained by quantum measurement [Eq.~(\ref{eq-P})].

Obviously, the complexity of $|\Psi\rangle$ scales exponentially with the number of spins, i.e., with the number of features $M$ while using the QFM given by Eq.~(\ref{eq-QFM}). The use of TN to represent $|\Psi\rangle$ lowers the complexity from exponential to polynomial. The TN state that correctly represents the joint probability distribution is referred to as GTN.

Below, we take MPS as the representation of GTN, which is written as
\begin{eqnarray}
	|\Psi\rangle =&& \sum_{s_1...s_M} \sum_{a_1...a_{M-1}} A^{[1]}_{s_1a_1} A^{[2]}_{s_2a_1a_2} ... \nonumber \\ &&  A^{[M-1]}_{s_{M-1}a_{M-2}a_{M-1}} A^{[M]}_{s_{M}a_{M-1}} |s_1\rangle...|s_M\rangle,
	\label{eq-MPS}
\end{eqnarray}
with $s_m=0,1$ and thus $\dim(s_m)=d=2$ (called physical dimension). The indexes $\{a_m\}$ are called virtual indexes. We take their dimensions to be uniformly $\dim(a_m) = \chi$, which is dubbed as virtual dimension. The complexity of an MPS just scales linearly with the number of spins $M$ as
\begin{eqnarray}
	\#|\Psi\rangle \sim O(Md\chi^2).
	\label{eq-complexity}
\end{eqnarray}

To train $|\Psi\rangle$ so that Eq.~(\ref{eq-P}) can correctly give the joint probability distribution, one can update the local tensors $\{\boldsymbol{A}^{[m]}\}$ of MPS to minimize the NLL
\begin{eqnarray}
	L = -\frac{1}{N} \sum_n \ln P(\boldsymbol{x}^{(n)}),
	\label{eq-nll}
\end{eqnarray}
with $\boldsymbol{x}^{(n)}$ denoting the $n$-th sample in a provided training set. Taking the gradient descent method as an example, the local tensors of MPS are updated as
\begin{eqnarray}
	\boldsymbol{A}^{[m]} \leftarrow \boldsymbol{A}^{[m]} - \eta \frac{\partial L}{\partial \boldsymbol{A}^{[m]}},
	\label{eq-gradient}
\end{eqnarray}
with $\eta$ a small positive number known as the gradient step or learning rate. 

\textit{Catastrophe of orthogonality.---} The COO of quantum many-body states says that the fidelity between two ``different'' quantum states generally decreases exponentially with the number of spins $M$ as
\begin{eqnarray}
	\left| \langle \Psi|\Phi \rangle \right| \sim k^M,
	\label{eq-fid}
\end{eqnarray}
with $0<k<1$. 

Below, let us focus on the COO of MPS's. We utilize the orthogonal form of MPS and place the orthogonal center on the right end. Then all tensors except the last one are isometries, satisfying the orthogonal conditions $\sum_{s_m a_{m-1}} A^{[m] \ast}_{s_ma_{m-1}a_m} A^{[m]}_{s_ma_{m-1}a'_m} = I_{a_m a'_m}$, with $\boldsymbol{I}$ denoting the identity. The rightmost tensor satisfies the normalization condition as $\sum_{s_M a_{M-1}} A^{[M] \ast}_{s_Ma_{M-1}} A^{[M]}_{s_Ma_{M-1}} = 1$, so that the whole MPS satisfies the normalization condition (say, $\left| \langle \Psi|\Phi \rangle \right|= 1$ for $|\Psi\rangle = e^{i\zeta} |\Phi\rangle$ with $\zeta$ an arbitrary phase factor). Note that any MPS can be transformed into such an orthogonal form by gauge transformations (meaning the transformations do not change the tensor obtained by contracting all virtual indices).

With $|\Psi\rangle$ and $|\Phi\rangle$ denoting two different MPS's (for instance, their local tensors, denoted as $\{\boldsymbol{A}^{[m]}\}$ and $\{\boldsymbol{B}^{[m]}\}$, are randomly generated), $\left| \langle \Psi|\Phi \rangle \right|$ would generally satisfy Eq.~(\ref{eq-fid}). To show this, we may compute $\left| \langle \Psi|\Phi \rangle \right|$ by contracting the shared indexes from left to right, starting with $v_{a_1 a'_1} = \sum_{s_1} A^{[1] \ast}_{s_1a_1} B^{[1]}_{s_1a'_1}$. As $\boldsymbol{A}^{[1]}$ and $\boldsymbol{B}^{[1]}$ are isometries, we have $|\boldsymbol{v}| \leq 1$, where the equal sign holds for $\boldsymbol{A}^{[1]} = \boldsymbol{B}^{[1]}$. We may normalize $\boldsymbol{v}$ as $\boldsymbol{v} \leftarrow \boldsymbol{v} / z_1$ with $z_1 = |\boldsymbol{v}| < 1$ (assuming $\boldsymbol{A}^{[1]} \neq \boldsymbol{B}^{[1]}$).

Subsequently, we contract $\boldsymbol{v}$ with $\boldsymbol{A}^{[m]}$ and $\boldsymbol{B}^{[m]}$ for $m=2, 3, \dots, M$. After each contraction, we will generally have $|\boldsymbol{v}| < 1$ as $\boldsymbol{A}^{[m]}$ and $\boldsymbol{B}^{[m]}$ are different isometries. Then we normalize it as $\boldsymbol{v} \leftarrow \boldsymbol{v} / z_m$ with $z_m = |\boldsymbol{v}| < 1$. Finally, we have $\left| \langle \Psi|\Phi \rangle \right| = \tilde{z}\prod_{m=1}^{M-1} z_m$ with $\tilde{z} = \left| \sum v_{a_{M-1} a'_{M-1}} A^{[M] \ast}_{s_Ma_{M-1}} B^{[M]}_{s_Ma'_{M-1}} \right|$, which is also generally less than $1$. Obviously, $\left| \langle \Psi|\Phi \rangle \right| \to 0$ exponentially with $M$, since it is the product of $M$ numbers that are generally less than $1$.

More rigorous deductions about the COO can be given if we impose certain restrictions on the MPS's, such as translational invariance ($\boldsymbol{A}^{[m]} = \boldsymbol{A}$ and $\boldsymbol{B}^{[m]} = \boldsymbol{B}$ for any $m$). The exponential decay of $\left| \langle \Psi|\Phi \rangle \right|$ can be immediately seen by the fact that the dominant eigenvalue of the transfer matrix ($M_{aa',bb'} = \sum_{s} A_{sab} B_{sa'b'}$) is in general less than $1$. This corresponds to the ``shrinkage'' of norm in the contractions discussed above. Considering that the GTN's are obtained by updating the local tensors with ML data, we will not further seek for more rigorous deductions, but will assume the satisfaction of COO with GTN's.

\begin{figure}[tbp]
	\centering
	\includegraphics[angle=0,width=0.75\linewidth]{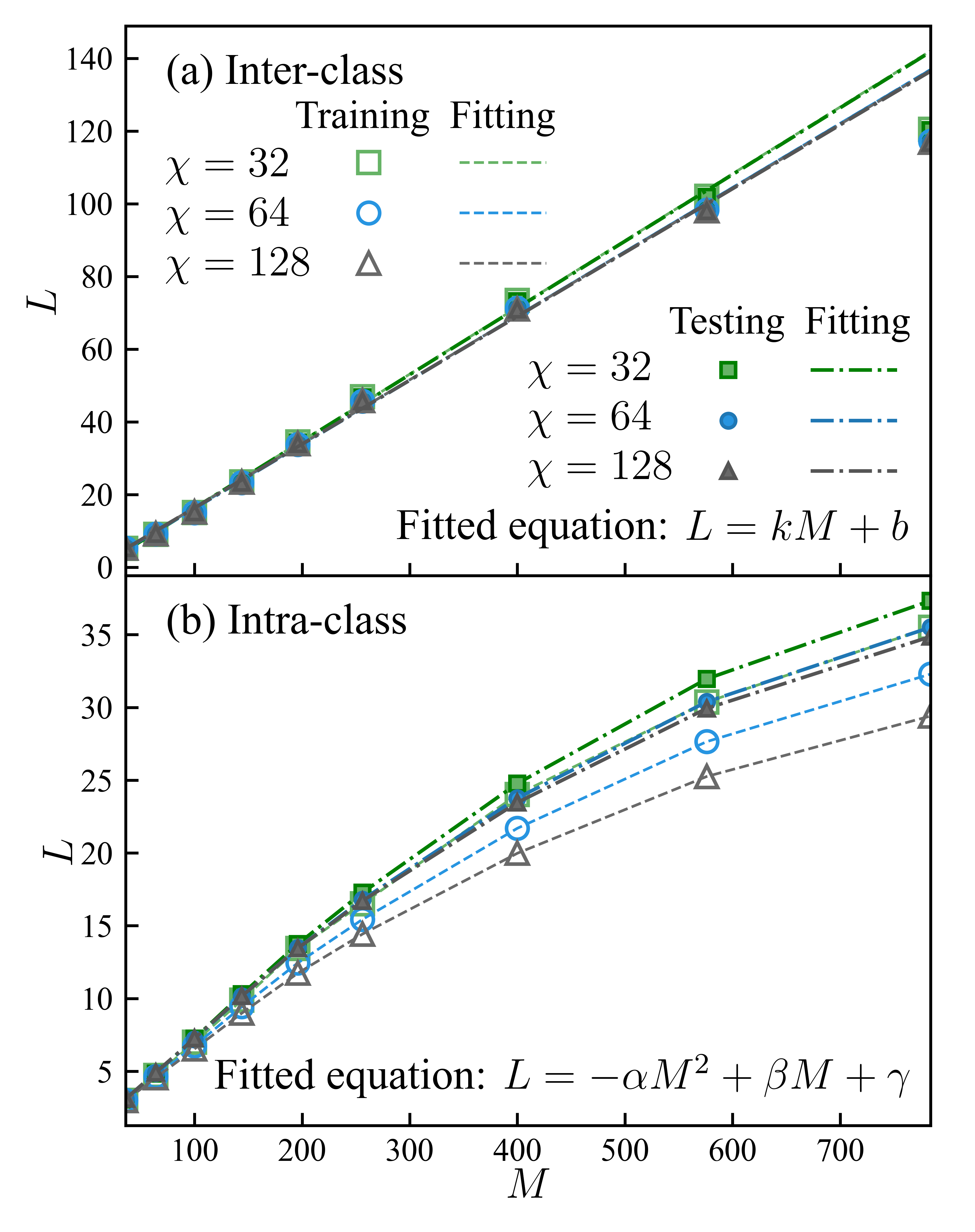} 
	\caption{(Color online) (a) The linear scaling of the ``inter-class'' NLL [Eq.~(\ref{eq-linear})], where the samples and GTN's correspond to different categories while computing NLL. (b) The emergence of the negative quadratic correction in the scaling of the ``intra-class'' NLL [Eq.~(\ref{eq-corrected})], where the samples and GTN's correspond to the same category. Here, we choose the Fashion-MNIST dataset. The data for $M<784$ are obtained by cropping the middle section of the images.}
	\label{fig-grayscale_theory}
\end{figure}

\textit{Universal scaling laws of negative logarithmic likelihood.---} From Eqs.~(\ref{eq-P}) and (\ref{eq-fid}), the probability of obtaining a sample ($\boldsymbol{x}$) from an irrelevant GTN should be exponentially small. Substituting this into Eq.~(\ref{eq-nll}), one immediately obtains the linear scaling law of NLL as given by Eq.~(\ref{eq-linear}). Fig.~\ref{fig-grayscale_theory}(a) demonstrates such a scaling law on the Fashion-MNIST dataset~\cite{fMNIST} for the inter-class NLL, where we take the categories of the samples to differ from those of the (trained) GTN's. By linear fitting, we have $k = 0.1837$, $0.1764$, and $0.1758$ for the training set ($k = 0.1835$, $0.1762$, and $0.1756$ for the testing set) for $\chi = 32$, $64$, and $128$, respectively. This indicates the scaling law of NLL does not change by varying $\chi$.

Fig.~\ref{fig-grayscale_theory}(b) shows that the intra-class NLL, where the categories of the samples should correspond to those of the GTN's, obeys the corrected scaling law given by Eq.~(\ref{eq-corrected}). We have the quadratic coefficient $\alpha = 3.653 \times 10^{-5}$, $3.252 \times 10^{-5}$, and $3.041 \times 10^{-5}$ for the training set ($\alpha = 3.663 \times 10^{-5}$, $3.582 \times 10^{-5}$, and $3.603 \times 10^{-5}$ for the testing set) for $\chi = 32$, $64$, and $128$, respectively. As $M \sim O(10^{2})$, we have $M^{2} \sim O(10^{4})$ or $O(10^{5})$. Thus, the contribution of the quadratic term is about $O(1)$, which is ignorable particularly for large $M$. For the linear term, we have $\beta = 7.365 \times 10^{-2}$, $6.582 \times 10^{-2}$, and $6.006 \times 10^{-2}$ for the training set ($\beta = 7.617 \times 10^{-2}$, $7.277 \times 10^{-2}$, and $7.196 \times 10^{-2}$ for the testing set). The magnitudes of constant terms are about $O(1)$ or $O(10^{-1})$. These indicate the contributions from all three terms cannot be ignored. The universality of the scaling laws is further demonstrated on various datasets~\cite{SM}.

\begin{figure}[tbp]
\centering
\includegraphics[angle=0,width=1\linewidth]{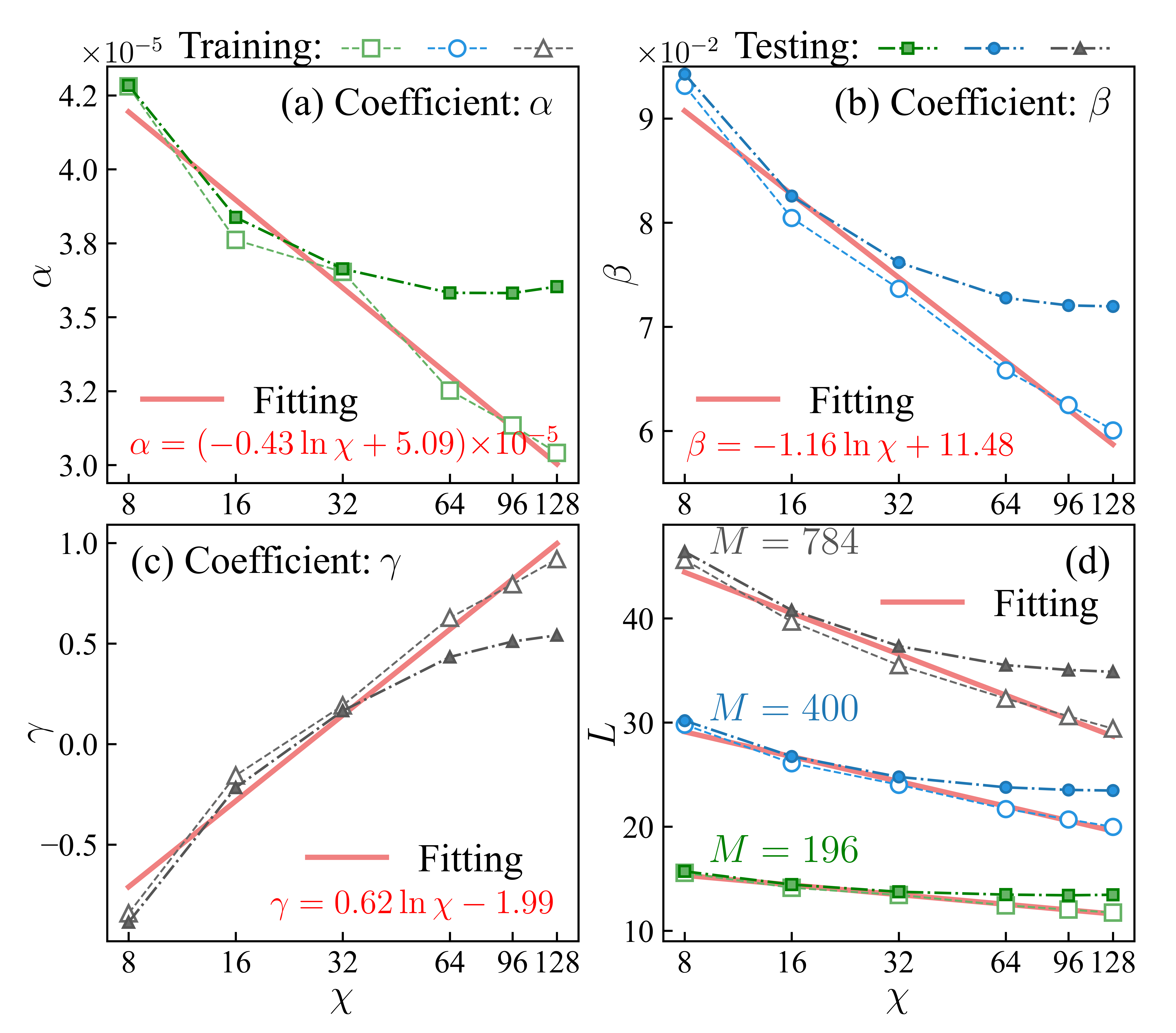}
\caption{(Color online) (a)-(c) The coefficients in the corrected scaling laws [$\alpha$, $\beta$, and $\gamma$ in Eq.~(\ref{eq-corrected})] of the intra-class NLL and their logarithmic scaling against the virtual dimension $\chi$. (c) The intra-class NLL $L$ and its logarithmic scaling against $\chi$. The deviations between the curves of the training and testing sets are observed, which indicate over-parameterization. We take the Fashion-MNIST dataset and fix $M=784$. The fittings (red solid lines) are given based on the training data.}
\label{fig-classification_binary}
\end{figure}

Below, we focus on the intra-class NLL, if not mentioned specifically. In Fig.~\ref{fig-classification_binary}(a), we change the virtual dimension $\chi$ of the GTN by fixing $M=784$ and show the values of $\alpha$ in Eq.~(\ref{eq-corrected}) for the intra-class NLL. Our results suggest the logarithmic scaling behavior of $\alpha$ against $\chi$ as
\begin{eqnarray}
	\alpha \simeq p_{\alpha} \ln \chi + q_{\alpha}.
	\label{eq-alpha}
\end{eqnarray}
Similar logarithmic behaviors are observed for the coefficients $\beta$ and $\gamma$, as shown in Fig.~\ref{fig-classification_binary}(b) and (c). 

Substituting such logarithmic relations of $\alpha$, $\beta$, and $\gamma$ [one may refer to Eq.~(\ref{eq-alpha})] to the corrected scaling law, we have
\begin{equation}
	L \simeq (-p_{\alpha} M^{2} + p_{\beta} M + p_{\gamma}) \ln \chi + (-q_{\alpha} M^{2} + q_{\beta} M + q_{\gamma}).
	\label{eq-Lchi}
\end{equation}
This suggests the logarithmic scaling of $L$ against $\chi$ as
\begin{equation}
	L \simeq p_{L}\ln \chi + q_{L},
	\label{eq-Llog}
\end{equation}
which is verified in Fig.~\ref{fig-classification_binary}(d). It means logarithmic improvement of the ML powers versus the number of quantum channels (which will be discussed below).

Another important observation in Fig.~\ref{fig-classification_binary}(a)-(c) is the deviation between the curves for the training and testing sets. For about $\chi > 30$, the coefficients in the corrected scaling law for the training set maintain the logarithmic scaling behavior given by Eq.~(\ref{eq-Llog}). But those for the testing set start to deviate, leading to higher NLL as shown in Fig.~\ref{fig-classification_binary}(d). These results imply that by further increasing the parameter complexity of GTN for about $\chi > 30$, the representation power improves by obeying the expected scaling law, but the improvement of generalization power decelerates. This can be regarded as a sign of over-parameterization. Notably, such a sign can be clearly observed with the deviation of the scaling of coefficients but can hardly be seen in NLL, which is one of the advantages of gaining the knowledge about the scaling laws.

\begin{figure}[tbp]
	\centering
	\includegraphics[angle=0,width=1\linewidth]{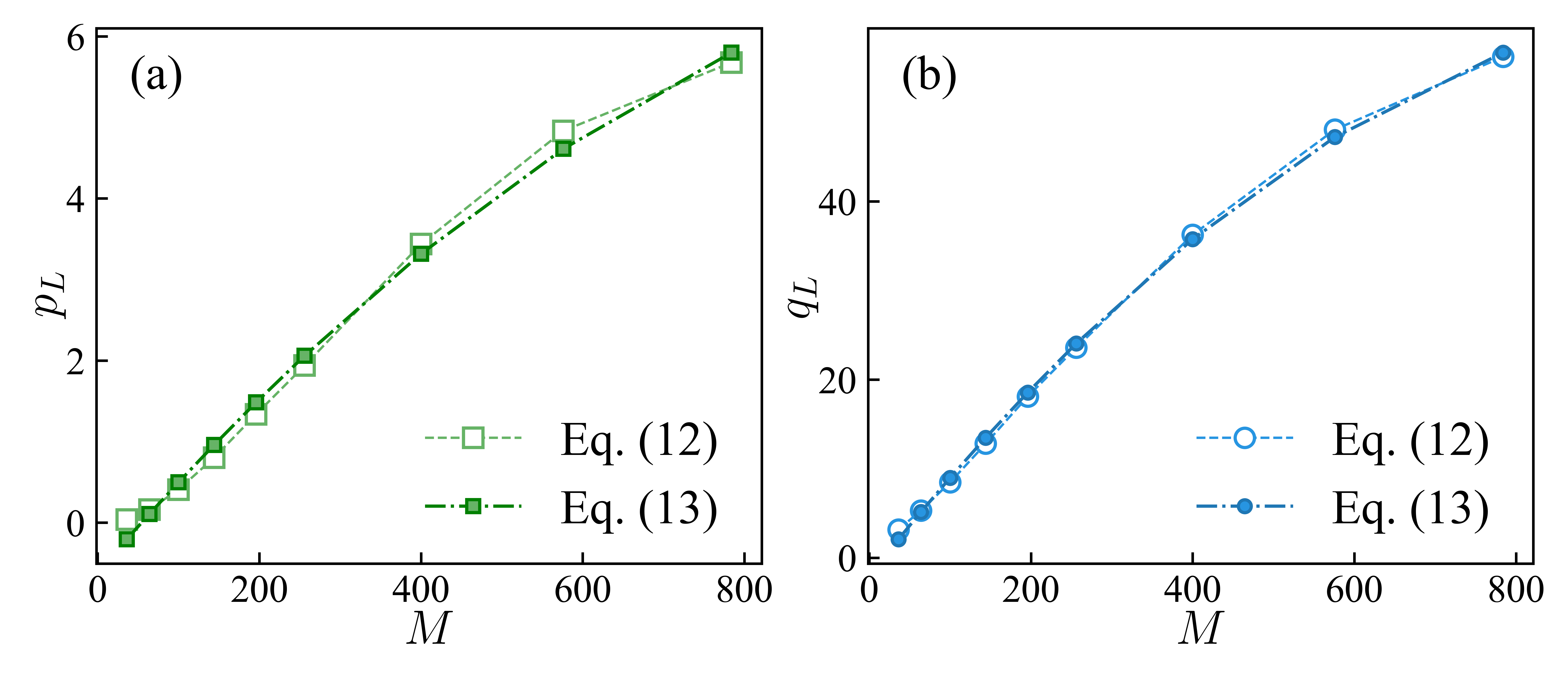}
	\caption{(Color online) Comparisons of (a) $p_L$ and (b) $q_L$ between the values obtained by the fitting with Eq.~(\ref{eq-Llog}) and those computed using Eq.~(\ref{eq-pq}).}
	\label{fig-fitfit}
\end{figure}

Combining Eqs.~(\ref{eq-Lchi}) and (\ref{eq-Llog}), we can immediately obtain the relations between the relevant coefficients as
\begin{equation}
	\begin{split}
		p_{L} &= -p_{\alpha} M^{2} + p_{\beta} M + p_{\gamma}, \\ 
		q_{L} &= -q_{\alpha} M^{2} + q_{\beta} M + q_{\gamma}.
	\end{split}
	\label{eq-pq}
\end{equation}
Fig.~\ref{fig-fitfit} verifies such relations by comparing the $p_{L}$ and $q_{L}$ obtained by Eq.~(\ref{eq-pq}) with those by fitting the data with Eq.~(\ref{eq-Llog}), which show remarkable coincidence.

\textit{Orthogonality, quantum probabilistic interpretation, and their relations to generalization power.---} To gain a deeper understanding of the representation and generalization powers of GTN, below we investigate the orthogonality of QFM (referring to the orthogonality of the states obtained by QFM of Eq.~(\ref{eq-QFM}) and the satisfaction of quantum probabilistic interpretation.

\begin{figure}[tbp]
	\centering
	\includegraphics[angle=0,width=1\linewidth]{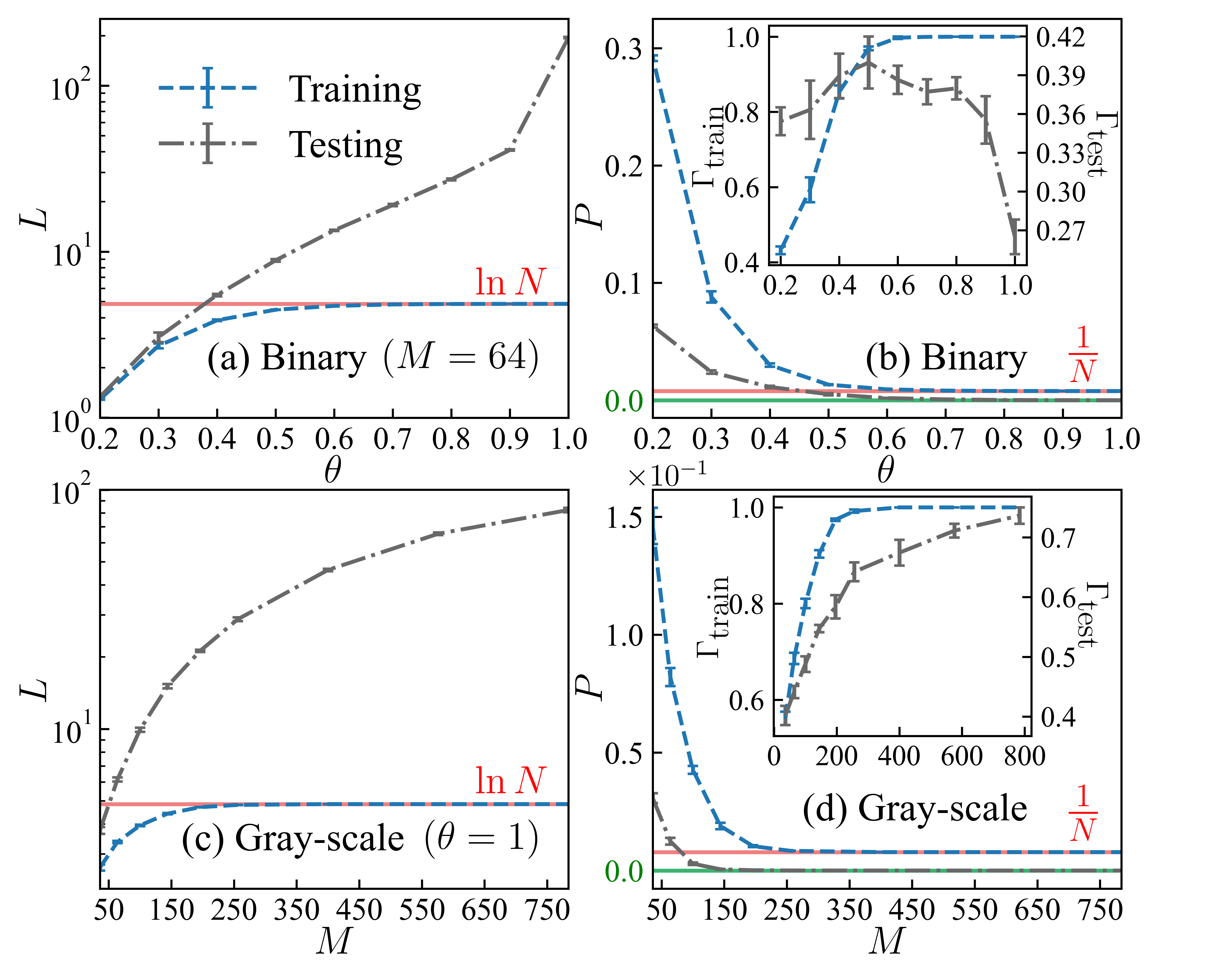} 
	\caption{(Color online) (a) The NLL $L$ and (b) average probability per sample $P$ for the training and testing sets of binarized Fashion-MNIST versus $\theta$ in the QFM [Eq.~(\ref{eq-QFM})]. We fix $M=64$. (c) and (d) show $L$ and $P$ versus $M$ for the original (gray-scale) dataset with $\theta=1$. In the insets of (b) and (d), we show the classification accuracy of the GTN classifier against $\theta$ and $M$, respectively. The error bars indicate the standard deviations over 5 independent simulations. We take $N = \chi = 128$.}
	\label{fig-generate_theta_M}
\end{figure}

Consider a simple case of binary images with $\chi = N$ (with $N$ the number of training samples), which can provide fundamental information on the quantum probabilistic ML with small $N$ or sufficiently-large $\chi$. Taking $\theta=1$ in Eq.~(\ref{eq-QFM}), we exactly have the orthogonality $\langle x|y \rangle = 0$ for any two distinct samples $x \neq y$. With a well-trained or constructed GTN, say $|\Psi\rangle = \frac{1}{\sqrt{N}} \sum_{n=1}^N |\boldsymbol{x}^{(n)}\rangle$ (that can be exactly written as an MPS with $\chi=N$), we have $P(\boldsymbol{x}^{(n)}) = 1/N$ for any training sample. 

For any samples (say a testing sample denoted as $\boldsymbol{y}$) that do not belong to the training set, we have $P(\boldsymbol{y}) = 0$ due to the orthogonality. One may understand $\chi$ as the number of channels in the GTN. $N$ orthogonal states have to occupy $N$ quantum channels. For $\chi=N$, all quantum channels in the GTN are taken by the training samples, which gives the ideal equal distribution of probabilities and minimal NLL $L = \ln N$. No channel is left for the testing samples, giving vanishing probabilities and divergent NLL. This is demonstrated in Fig.~\ref{fig-generate_theta_M}(a) and (b) with the binarized Fashion-MNIST dataset.

In this binary case with $\theta=1$ above, the normalization condition for the full-Hilbert-space quantum probabilities is satisfied, i.e., $\sum_{\boldsymbol{x}} P(\boldsymbol{x}) = 1$ (since $\sum_{\boldsymbol{x}} P(\boldsymbol{x}) = \langle \Psi|\Psi \rangle$) with the summation through all possible samples. Thus, the obtained NLL is always larger than the theoretical minimum $\ln N$. But for $0<\theta<1$, the product states obtained by QFM from different samples are generally not orthogonal to each other. Consequently, reducing $\theta$ will cause an ``impertinent'' decrease of NLL, essentially due to the violation of the normalization condition of quantum probabilities, i.e., $\sum_{\boldsymbol{x}} P(\boldsymbol{x}) > 1$. Consequently, the NLL can become lower than its theoretical minimum $L < \ln N$ [see Fig.~\ref{fig-generate_theta_M}(a)].

The NLL for the testing set also decreases with $\theta$, which is also ``impertinent'' but might be a positive signal on the generalization power (we will show this later). Fig.~\ref{fig-generate_theta_M}(b) shows the unusually large average probability $P$ for both the training and testing sets while reducing $\theta \to 0$.

Now we adopt the GTN classification (GTNC) scheme~\cite{sun2020generative} to demonstrate the representation and generalization powers from another perspective. The idea is to use $G$ GTN's (with $G$ the number of categories) to parameterize the classification boundaries of samples in the Hilbert space of quantum states (the GTN's). Given a sample $\boldsymbol{y}$, the GTNC predicts its classification as 
\begin{eqnarray}
	g^{\text{pred}}(\boldsymbol{y}) = \text{argmax}_g \left| \langle \boldsymbol{y}| \Psi^{(g)} \rangle \right|,
	\label{eq-gtnc}
\end{eqnarray}
with $| \Psi^{(g)} \rangle$ the GTN trained by the training samples in the $g$-th category.

The inset of Fig.~\ref{fig-generate_theta_M}(b) shows the classification accuracy against $\theta$ for the training and testing sets of the binarized Fashion-MNIST dataset. For $\theta=1$, the training accuracy is theoretically $100\%$, while the testing accuracy is low. This is a typical example of over-parameterization, which can be the result of insufficient training data or redundant parameters (note that here we take $N=\chi$).

By reducing $\theta$, the normalization condition of the full-Hilbert-space quantum probabilities is violated, while the training accuracy decreases with $\theta$. However, non-vanishing probabilities appear for the testing samples, which improve the testing accuracy (see the peak of testing accuracy around $\theta=0.5$). These results imply a balance between the violation of the quantum normalization condition and the generalization power of quantum states for ML, tuned by the orthogonality of QFM.

The above implication is meaningful for handling non-binary data, which are encountered more frequently than the binary ones. For instance, each feature (pixel) in an 8-bit gray-scale image can take $D=256$ distinct values. If we rigorously require orthogonality such that $\langle x|y \rangle = 0$ for $x \neq y$, we should map each pixel to a spin with $D$ levels, which is infeasible and impractical. Our results provide an interpretation of why the QFM with two-level ($d=2$) spins works well on gray-scale images, though it clearly leads to the violation of the quantum normalization condition.

In Fig.~\ref{fig-generate_theta_M}(c) and (d), we show the NLL $L$ and average probability $P$ against $M$ for the original (gray-scale) Fashion-MNIST dataset. For $\theta=1$, we still do not have orthogonality; generally, $|\langle x|y \rangle| > 0$ for $x \neq y$. This explains why the NLL of the training set is lower than the theoretical minimum $\ln N$. However, we have $L \to \ln N$ and $P \to 1/N$ for large $M$, suggesting the restoration of orthogonality (and the normalization condition) thanks to COO. In the inset of Fig.~\ref{fig-generate_theta_M}(d), we show that the testing accuracy increases with $M$, implying that for $D \ll d$, enhancing orthogonality will generally improve the generalization power~\cite{SM}.

\textit{Summary.---} The main contribution of this work is the proposal of universal scaling laws in quantum probabilistic machine learning, where the joint probabilistic distributions are encoded as generative tensor networks in the form of matrix product states. We show that the gain of representation or generalization power suppresses the linear scaling law of negative logarithmic likelihood, which is a result of the catastrophe of orthogonality of quantum many-body states, by introducing a negative quadratic correction. The relations among the scaling laws, orthogonality of the quantum feature map, and the normalization condition of quantum probabilities are discussed, providing a systematic understanding of the representation and generalization powers of quantum probabilistic machine learning.

\section*{Acknowledgment} 
SCB is grateful to Yi-Cheng Tang, Rui Hong, Peng-Fei Zhou, and Ying Lu for helpful discussions. This work was supported in part by Beijing Natural Science Foundation (Grant No. 1232025) and Academy for Multidisciplinary Studies, Capital Normal University.

%
%

\FloatBarrier 
\appendix

\renewcommand\thefigure{A\arabic{figure}}
\renewcommand\theequation{A\arabic{equation}}
\renewcommand\thetable{A\arabic{table}}
\setcounter{figure}{0}
\setcounter{equation}{0}
\setcounter{table}{0}

\section{Appendix I. Universality of scaling laws}

\begin{figure}[tbp]
	\centering
	\includegraphics[angle=0,width=0.75\linewidth]{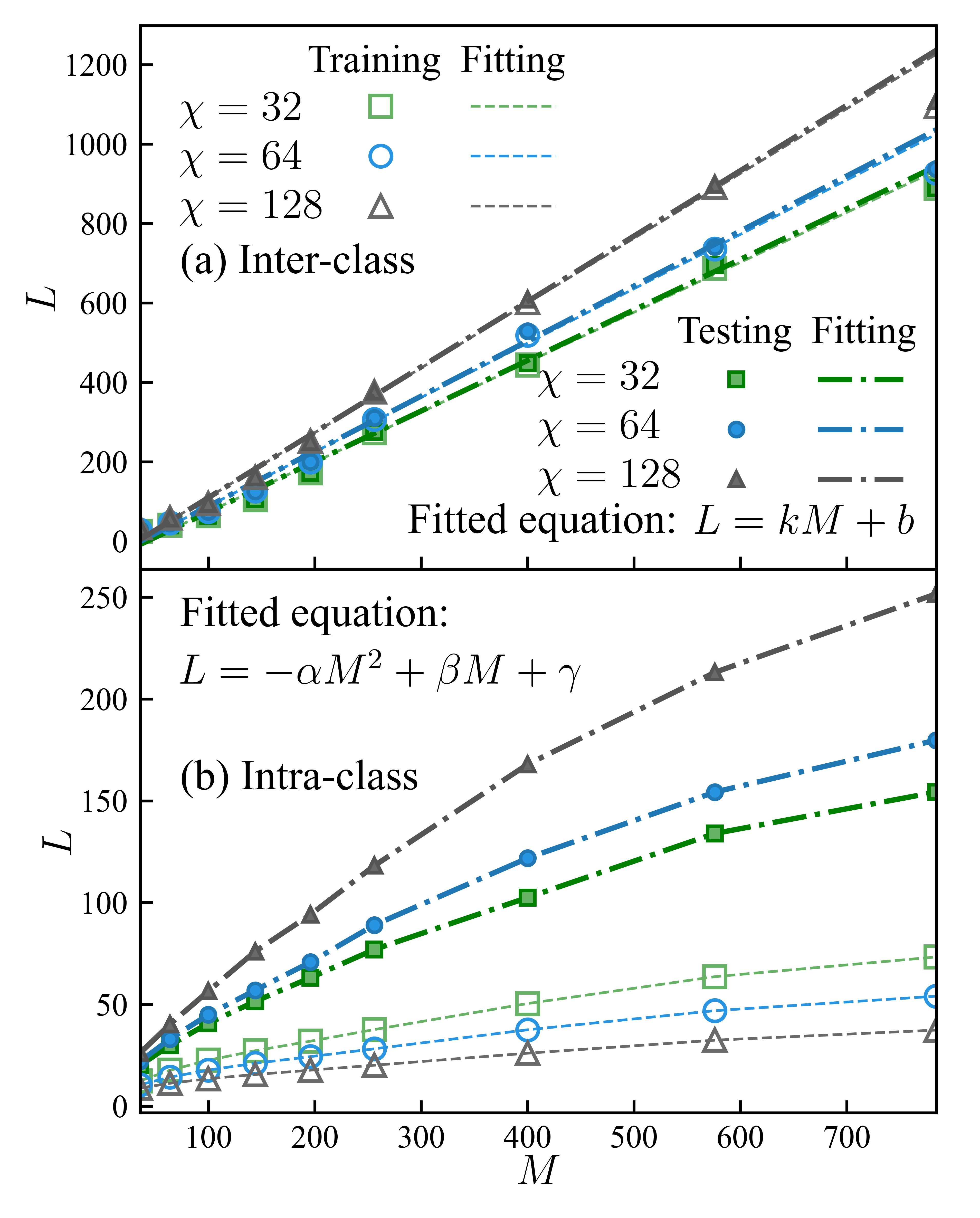} 
	\caption{(Color online) (a) The linear scaling of the ``inter-class'' NLL and (b) the emergence of the negative quadratic correction in the scaling of the ``intra-class'' NLL on the binarized Fashion-MNIST dataset. The data for $M<784$ are obtained by cropping the middle section of the images.}
	\label{fig-appendix_orthogonal_theory}
\end{figure}	

Below, we further demonstrate the universality of scaling laws on the binarized Fashion-MNIST dataset. Fig.~\ref{fig-appendix_orthogonal_theory}(a) shows the linear scaling law of the ``inter-class'' negative logarithmic likelihood (NLL) $L \simeq kM + b$, which is a consequence of the catastrophe of orthogonality. We have $k = 1.257$, $1.369$, and $1.631$ for the training set (and $k = 1.273$, $1.386$, and $1.647$ for the testing set) for $\chi = 32$, $64$, and $128$, respectively.

Fig.~\ref{fig-appendix_orthogonal_theory}(b) shows the scaling law of the ``inter-class'' NLL corrected with a quadratic term: $L \simeq \beta M - \alpha M^2 + \gamma$. We have the quadratic coefficient $\alpha = 5.827 \times 10^{-5}$, $3.981 \times 10^{-5}$, and $2.284 \times 10^{-5}$ for the training set (and $\alpha = 1.330 \times 10^{-4}$, $1.624 \times 10^{-4}$, and $2.191 \times 10^{-4}$ for the testing set) for $\chi = 32$, $64$, and $128$, respectively.

\begin{figure}[tbp]
	\centering
	\includegraphics[angle=0,width=0.75\linewidth]{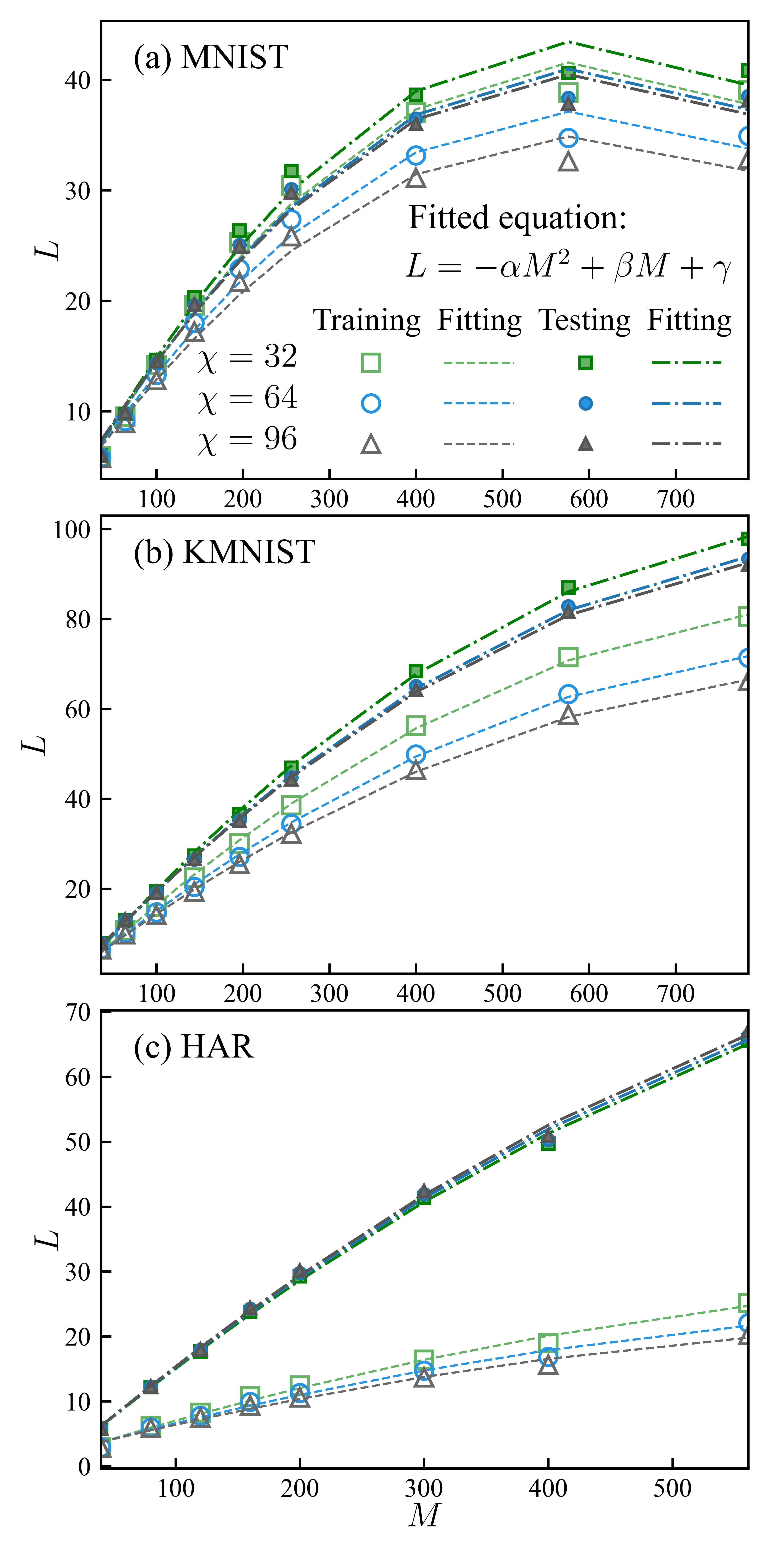} 
	\caption{(Color online) The scaling laws of the ``intra-class'' NLL for (a) MNIST~\cite{MNIST_6296535}, (b) Kuzushiji-MNIST (KMNIST)~\cite{clanuwat2018deep}, and (c) Human Activity Recognition (HAR)~\cite{human_activity_recognition_using_smartphones_240} datasets. The data for different $M$'s are obtained by cropping the middle section.}
	\label{fig-appendix_different datasets}
\end{figure}

Compared with the results from the original (gray-scale) Fashion-MNIST dataset, the NLL with the binary data is much higher due to the orthogonality of the QFM. The orthogonality can be measured by the fidelity $\left| \langle \boldsymbol{x}| \boldsymbol{y} \rangle \right|$. Lower fidelities mean larger quantum fluctuations in the GTN, thus leading to higher NLL. Nevertheless, the scaling laws robustly emerge. We also demonstrate such scaling laws in three different datasets (Fig.~\ref{fig-appendix_different datasets}), which are MNIST consisting of hand-drawn digits~\cite{MNIST_6296535}, Kuzushiji-MNIST (KMNIST), which consists of hand-written hiragana characters~\cite{clanuwat2018deep}, and Human Activity Recognition (HAR), which consists of acceleration data during human activities (such as walking, going upstairs, and going downstairs)~\cite{human_activity_recognition_using_smartphones_240}. Note that the data for different $M$'s are obtained by cropping the middle section. For MNIST, $L$ remains unchanged or slightly decreases for about $M>600$, possibly because the boundary regions of these images are almost blank.

\section{Appendix II. Scaling of negative logarithmic likelihood against number of training samples}

\begin{figure}[tbp]
	\centering
	\includegraphics[angle=0,width=0.8\linewidth]{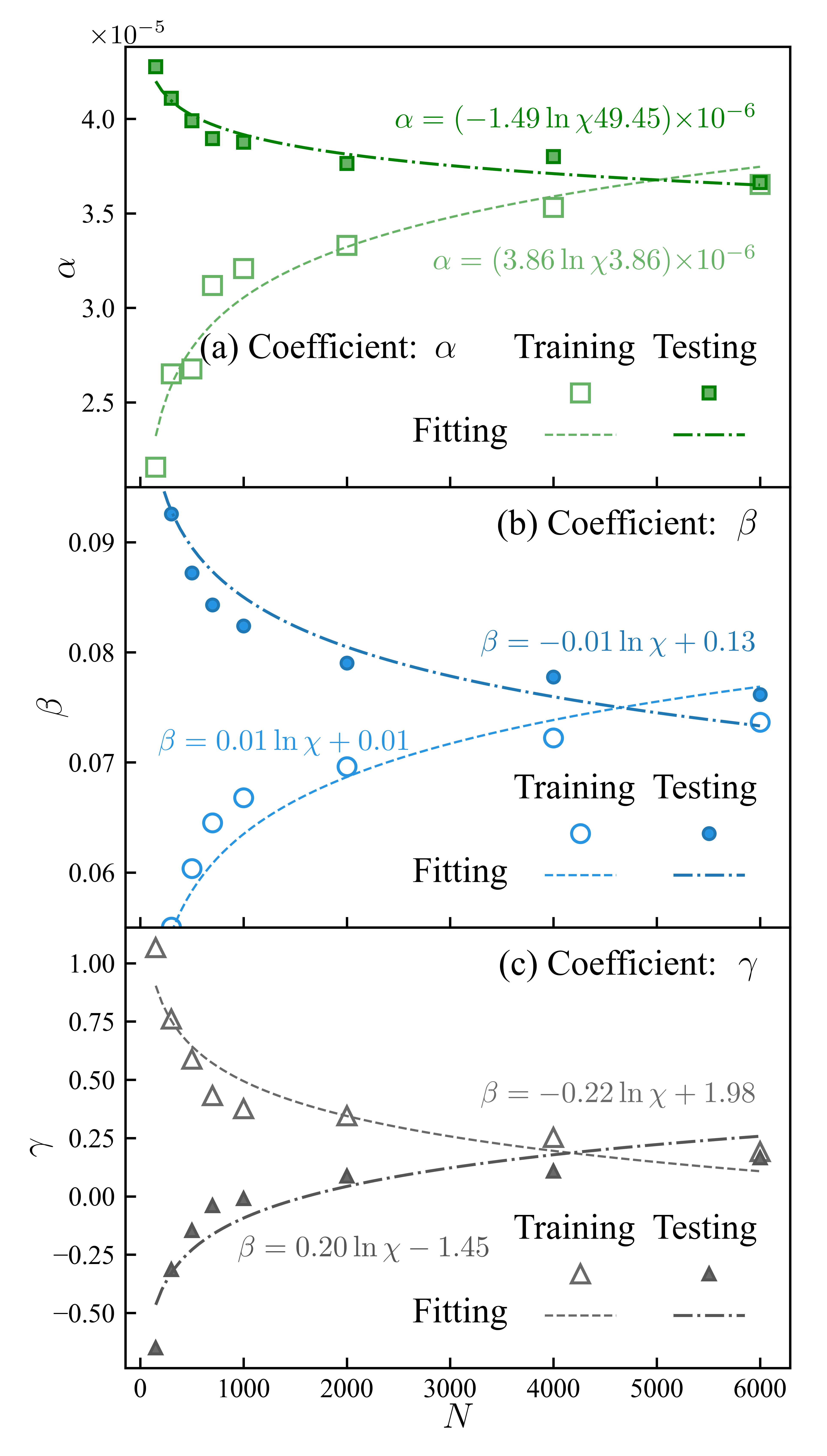} 
	\caption{(Color online) The logarithmic scaling behaviors if the coefficients (a) $\alpha$, (b) $\beta$, and (c) $\gamma$ in Eq.~(\ref{eq-corrected}) for the intra-class NLL scaling law for the Fashion-MNIST dataset against the number of training samples $N$.}
	\label{fig-appendix_grayscale_fitting_coefficient_with_N}
\end{figure}

In the main text, we show the utilization of scaling laws to identify over-parameterization while increasing the parameter complexity of the GTN. Fig.~\ref{fig-appendix_grayscale_fitting_coefficient_with_N} demonstrates the algorithmic scaling of NLL against the number of training samples $N$. The coefficients for the training and testing sets converge to the same values as $N$ increases. The discrepancy can be used to judge the sufficiency of the training data.

\section{Appendix III. Representation and generalization powers in tuning orthogonality}

\begin{figure}[tbp]
	\centering
	\includegraphics[angle=0,width=1\linewidth]{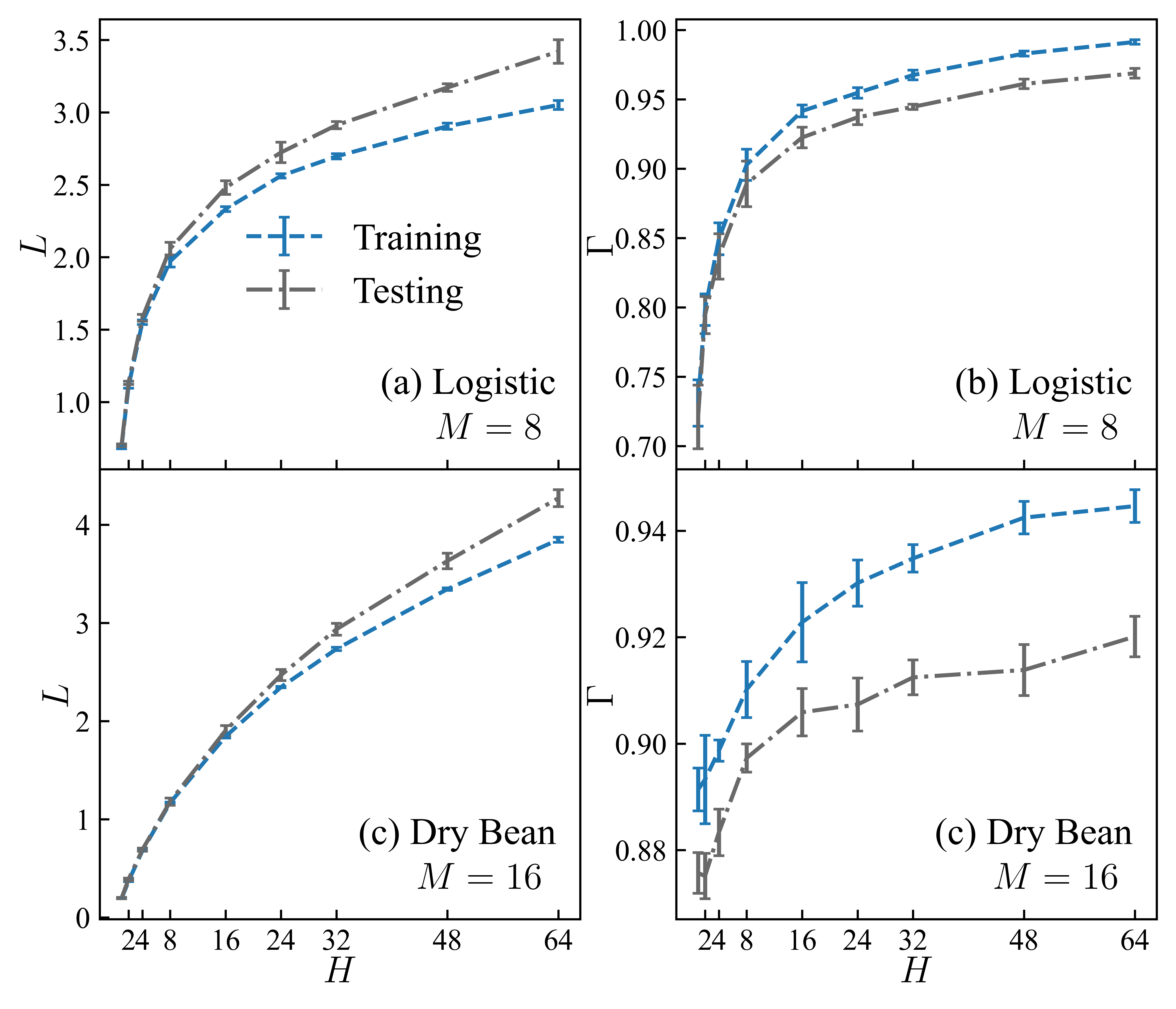}
	\caption{(Color online) (a) The NLL $L$ and (b) the classification accuracy $\Gamma$ by GTNC against $H$ in the msQFM [Eq.~(\ref{eq-msQFM})]on the datasets of Logistic dynamics~\cite{may1976simple} and dry bean~\cite{dry_bean_602}. The error bars represent the standard deviations over 5 independent simulations.}
	\label{fig-classification_chaos_beans}
\end{figure}

In the main text, we demonstrated how orthogonality affects the representation and generalization powers by adjusting the hyper-parameter $\theta$ in the quantum feature map (QFM) [Eq.~(\ref{eq-QFM})]. Below, we aim to adjust the orthogonality in a different manner to show how it affects the representation and generalization powers. 

To this aim, we define a multi-spin QFM (msQFM) $\boldsymbol{x} \to |\boldsymbol{x} \rangle = \prod_{m=1}^{M} |x_{m}\rangle$  with
\begin{equation}
	|x_m\rangle = \prod_{h=1}^H \left[ \cos \frac{\theta \pi x_m}{2} |0\rangle + \sin \frac{\theta \pi x_m}{2} |1\rangle \right],
	\label{eq-msQFM}
\end{equation}
with $H$ the number spins for encoding one feature. Therefore, the number of spins in a GTN significantly increases to $M_{\text{tot}} = MH$. Consequently, the orthogonality is enhanced due to the ``catastrophe of orthogonality (COO)'', particularly for the non-binary data. This is supported by the results in Fig.~\ref{fig-classification_chaos_beans}(a) and (b) on the datasets of Logistic dynamics~\cite{may1976simple} and dry bean~\cite{dry_bean_602}. Note that the Logistic dataset consists of the trajectories (time series) obtained by the evolutions of Logistic dynamics. Each trajectory represents a sample that is the result of $M$ steps of evolutions. The NLL $L$ increases with $H$, similar to the cases while changing $M$ (see, e.g., Figs. \ref{fig-appendix_orthogonal_theory}(b) and \ref{fig-appendix_different datasets}, as well as the results reported in the main text).

Increasing $H$ will not introduce any new information to the model. But, as shown in Fig.~\ref{fig-classification_chaos_beans}, the classification accuracy of the GTN classifier (GTNC)~\cite{sun2020generative} increases with $H$. This is consistent with the results given in the main text, indicating that for the non-binary data, enhancing orthogonality improves the representation and generalization powers. These results are particularly important for the non-binary datasets with relatively small $M$ (e.g., the Logistic and dry-bean datasets with $M=8$ and $16$, respectively, and many others), where the product states obtained by QFM are quite far away from being orthogonal even with $\theta=1$.
\normalem
%


\begin{thebibliography}{41}%
\makeatletter
\providecommand \@ifxundefined [1]{%
 \@ifx{#1\undefined}
}%
\providecommand \@ifnum [1]{%
 \ifnum #1\expandafter \@firstoftwo
 \else \expandafter \@secondoftwo
 \fi
}%
\providecommand \@ifx [1]{%
 \ifx #1\expandafter \@firstoftwo
 \else \expandafter \@secondoftwo
 \fi
}%
\providecommand \natexlab [1]{#1}%
\providecommand \enquote  [1]{``#1''}%
\providecommand \bibnamefont  [1]{#1}%
\providecommand \bibfnamefont [1]{#1}%
\providecommand \citenamefont [1]{#1}%
\providecommand \href@noop [0]{\@secondoftwo}%
\providecommand \href [0]{\begingroup \@sanitize@url \@href}%
\providecommand \@href[1]{\@@startlink{#1}\@@href}%
\providecommand \@@href[1]{\endgroup#1\@@endlink}%
\providecommand \@sanitize@url [0]{\catcode `\\12\catcode `\$12\catcode
  `\&12\catcode `\#12\catcode `\^12\catcode `\_12\catcode `\%12\relax}%
\providecommand \@@startlink[1]{}%
\providecommand \@@endlink[0]{}%
\providecommand \url  [0]{\begingroup\@sanitize@url \@url }%
\providecommand \@url [1]{\endgroup\@href {#1}{\urlprefix }}%
\providecommand \urlprefix  [0]{URL }%
\providecommand \Eprint [0]{\href }%
\providecommand \doibase [0]{https://doi.org/}%
\providecommand \selectlanguage [0]{\@gobble}%
\providecommand \bibinfo  [0]{\@secondoftwo}%
\providecommand \bibfield  [0]{\@secondoftwo}%
\providecommand \translation [1]{[#1]}%
\providecommand \BibitemOpen [0]{}%
\providecommand \bibitemStop [0]{}%
\providecommand \bibitemNoStop [0]{.\EOS\space}%
\providecommand \EOS [0]{\spacefactor3000\relax}%
\providecommand \BibitemShut  [1]{\csname bibitem#1\endcsname}%
\let\auto@bib@innerbib\@empty
\bibitem [{\citenamefont {Gilpin}\ \emph {et~al.}(2018)\citenamefont {Gilpin},
  \citenamefont {Bau}, \citenamefont {Yuan}, \citenamefont {Bajwa},
  \citenamefont {Specter},\ and\ \citenamefont {Kagal}}]{gilpin2018explaining}%
  \BibitemOpen
  \bibfield  {author} {\bibinfo {author} {\bibfnamefont {L.~H.}\ \bibnamefont
  {Gilpin}}, \bibinfo {author} {\bibfnamefont {D.}~\bibnamefont {Bau}},
  \bibinfo {author} {\bibfnamefont {B.~Z.}\ \bibnamefont {Yuan}}, \bibinfo
  {author} {\bibfnamefont {A.}~\bibnamefont {Bajwa}}, \bibinfo {author}
  {\bibfnamefont {M.}~\bibnamefont {Specter}},\ and\ \bibinfo {author}
  {\bibfnamefont {L.}~\bibnamefont {Kagal}},\ }\bibfield  {title} {\bibinfo
  {title} {Explaining explanations: An overview of interpretability of machine
  learning},\ }in\ \href {https://doi.org/10.1109/DSAA.2018.00018} {\emph
  {\bibinfo {booktitle} {2018 IEEE 5th International Conference on data science
  and advanced analytics (DSAA)}}}\ (\bibinfo {organization} {IEEE},\ \bibinfo
  {year} {2018})\ pp.\ \bibinfo {pages} {80--89}\BibitemShut {NoStop}%
\bibitem [{\citenamefont {Zhang}\ and\ \citenamefont
  {Zhu}(2018)}]{zhang2018visual}%
  \BibitemOpen
  \bibfield  {author} {\bibinfo {author} {\bibfnamefont {Q.-s.}\ \bibnamefont
  {Zhang}}\ and\ \bibinfo {author} {\bibfnamefont {S.-C.}\ \bibnamefont
  {Zhu}},\ }\bibfield  {title} {\bibinfo {title} {Visual interpretability for
  deep learning: a survey},\ }\href {https://doi.org/10.1631/FITEE.1700808}
  {\bibfield  {journal} {\bibinfo  {journal} {Frontiers of Information
  Technology \& Electronic Engineering}\ }\textbf {\bibinfo {volume} {19}},\
  \bibinfo {pages} {27} (\bibinfo {year} {2018})}\BibitemShut {NoStop}%
\bibitem [{\citenamefont {Carvalho}\ \emph {et~al.}(2019)\citenamefont
  {Carvalho}, \citenamefont {Pereira},\ and\ \citenamefont
  {Cardoso}}]{carvalho2019machine}%
  \BibitemOpen
  \bibfield  {author} {\bibinfo {author} {\bibfnamefont {D.~V.}\ \bibnamefont
  {Carvalho}}, \bibinfo {author} {\bibfnamefont {E.~M.}\ \bibnamefont
  {Pereira}},\ and\ \bibinfo {author} {\bibfnamefont {J.~S.}\ \bibnamefont
  {Cardoso}},\ }\bibfield  {title} {\bibinfo {title} {Machine learning
  interpretability: A survey on methods and metrics},\ }\href
  {https://doi.org/10.3390/electronics8080832} {\bibfield  {journal} {\bibinfo
  {journal} {Electronics}\ }\textbf {\bibinfo {volume} {8}},\ \bibinfo {pages}
  {832} (\bibinfo {year} {2019})}\BibitemShut {NoStop}%
\bibitem [{\citenamefont {Stoudenmire}\ and\ \citenamefont
  {Schwab}(2016)}]{SS16TNML}%
  \BibitemOpen
  \bibfield  {author} {\bibinfo {author} {\bibfnamefont {E.}~\bibnamefont
  {Stoudenmire}}\ and\ \bibinfo {author} {\bibfnamefont {D.~J.}\ \bibnamefont
  {Schwab}},\ }\bibfield  {title} {\bibinfo {title} {Supervised learning with
  tensor networks},\ }in\ \href
  {https://proceedings.neurips.cc/paper/2016/file/5314b9674c86e3f9d1ba25ef9bb32895-Paper.pdf}
  {\emph {\bibinfo {booktitle} {Advances in Neural Information Processing
  Systems}}},\ Vol.~\bibinfo {volume} {29},\ \bibinfo {editor} {edited by\
  \bibinfo {editor} {\bibfnamefont {D.}~\bibnamefont {Lee}}, \bibinfo {editor}
  {\bibfnamefont {M.}~\bibnamefont {Sugiyama}}, \bibinfo {editor}
  {\bibfnamefont {U.}~\bibnamefont {Luxburg}}, \bibinfo {editor} {\bibfnamefont
  {I.}~\bibnamefont {Guyon}},\ and\ \bibinfo {editor} {\bibfnamefont
  {R.}~\bibnamefont {Garnett}}}\ (\bibinfo  {publisher} {Curran Associates,
  Inc.},\ \bibinfo {year} {2016})\BibitemShut {NoStop}%
\bibitem [{\citenamefont {Liu}\ \emph {et~al.}(2019)\citenamefont {Liu},
  \citenamefont {Ran}, \citenamefont {Wittek}, \citenamefont {Peng},
  \citenamefont {Garc{\'{\i}}a}, \citenamefont {Su},\ and\ \citenamefont
  {Lewenstein}}]{liu2019machine}%
  \BibitemOpen
  \bibfield  {author} {\bibinfo {author} {\bibfnamefont {D.}~\bibnamefont
  {Liu}}, \bibinfo {author} {\bibfnamefont {S.-J.}\ \bibnamefont {Ran}},
  \bibinfo {author} {\bibfnamefont {P.}~\bibnamefont {Wittek}}, \bibinfo
  {author} {\bibfnamefont {C.}~\bibnamefont {Peng}}, \bibinfo {author}
  {\bibfnamefont {R.~B.}\ \bibnamefont {Garc{\'{\i}}a}}, \bibinfo {author}
  {\bibfnamefont {G.}~\bibnamefont {Su}},\ and\ \bibinfo {author}
  {\bibfnamefont {M.}~\bibnamefont {Lewenstein}},\ }\bibfield  {title}
  {\bibinfo {title} {Machine learning by unitary tensor network of hierarchical
  tree structure},\ }\href {https://doi.org/10.1088/1367-2630/ab31ef}
  {\bibfield  {journal} {\bibinfo  {journal} {New Journal of Physics}\ }\textbf
  {\bibinfo {volume} {21}},\ \bibinfo {pages} {073059} (\bibinfo {year}
  {2019})}\BibitemShut {NoStop}%
\bibitem [{\citenamefont {Han}\ \emph {et~al.}(2018)\citenamefont {Han},
  \citenamefont {Wang}, \citenamefont {Fan}, \citenamefont {Wang},\ and\
  \citenamefont {Zhang}}]{han2018unsupervised}%
  \BibitemOpen
  \bibfield  {author} {\bibinfo {author} {\bibfnamefont {Z.-Y.}\ \bibnamefont
  {Han}}, \bibinfo {author} {\bibfnamefont {J.}~\bibnamefont {Wang}}, \bibinfo
  {author} {\bibfnamefont {H.}~\bibnamefont {Fan}}, \bibinfo {author}
  {\bibfnamefont {L.}~\bibnamefont {Wang}},\ and\ \bibinfo {author}
  {\bibfnamefont {P.}~\bibnamefont {Zhang}},\ }\bibfield  {title} {\bibinfo
  {title} {Unsupervised generative modeling using matrix product states},\
  }\href {https://doi.org/10.1103/PhysRevX.8.031012} {\bibfield  {journal}
  {\bibinfo  {journal} {Phys. Rev. X}\ }\textbf {\bibinfo {volume} {8}},\
  \bibinfo {pages} {031012} (\bibinfo {year} {2018})}\BibitemShut {NoStop}%
\bibitem [{\citenamefont {Cheng}\ \emph {et~al.}(2019)\citenamefont {Cheng},
  \citenamefont {Wang}, \citenamefont {Xiang},\ and\ \citenamefont
  {Zhang}}]{cheng2019tree}%
  \BibitemOpen
  \bibfield  {author} {\bibinfo {author} {\bibfnamefont {S.}~\bibnamefont
  {Cheng}}, \bibinfo {author} {\bibfnamefont {L.}~\bibnamefont {Wang}},
  \bibinfo {author} {\bibfnamefont {T.}~\bibnamefont {Xiang}},\ and\ \bibinfo
  {author} {\bibfnamefont {P.}~\bibnamefont {Zhang}},\ }\bibfield  {title}
  {\bibinfo {title} {Tree tensor networks for generative modeling},\ }\href
  {https://doi.org/10.1103/PhysRevB.99.155131} {\bibfield  {journal} {\bibinfo
  {journal} {Phys. Rev. B}\ }\textbf {\bibinfo {volume} {99}},\ \bibinfo
  {pages} {155131} (\bibinfo {year} {2019})}\BibitemShut {NoStop}%
\bibitem [{\citenamefont {Sun}\ \emph {et~al.}(2020)\citenamefont {Sun},
  \citenamefont {Peng}, \citenamefont {Liu}, \citenamefont {Ran},\ and\
  \citenamefont {Su}}]{sun2020generative}%
  \BibitemOpen
  \bibfield  {author} {\bibinfo {author} {\bibfnamefont {Z.-Z.}\ \bibnamefont
  {Sun}}, \bibinfo {author} {\bibfnamefont {C.}~\bibnamefont {Peng}}, \bibinfo
  {author} {\bibfnamefont {D.}~\bibnamefont {Liu}}, \bibinfo {author}
  {\bibfnamefont {S.-J.}\ \bibnamefont {Ran}},\ and\ \bibinfo {author}
  {\bibfnamefont {G.}~\bibnamefont {Su}},\ }\bibfield  {title} {\bibinfo
  {title} {Generative tensor network classification model for supervised
  machine learning},\ }\href {https://doi.org/10.1103/PhysRevB.101.075135}
  {\bibfield  {journal} {\bibinfo  {journal} {Phys. Rev. B}\ }\textbf {\bibinfo
  {volume} {101}},\ \bibinfo {pages} {075135} (\bibinfo {year}
  {2020})}\BibitemShut {NoStop}%
\bibitem [{\citenamefont {Bai}\ \emph {et~al.}(2022)\citenamefont {Bai},
  \citenamefont {Tang},\ and\ \citenamefont {Ran}}]{Bai:100701}%
  \BibitemOpen
  \bibfield  {author} {\bibinfo {author} {\bibfnamefont {S.-C.}\ \bibnamefont
  {Bai}}, \bibinfo {author} {\bibfnamefont {Y.-C.}\ \bibnamefont {Tang}},\ and\
  \bibinfo {author} {\bibfnamefont {S.-J.}\ \bibnamefont {Ran}},\ }\bibfield
  {title} {\bibinfo {title} {Unsupervised recognition of informative features
  via tensor network machine learning and quantum entanglement variations},\
  }\href {https://doi.org/10.1088/0256-307X/39/10/100701} {\bibfield  {journal}
  {\bibinfo  {journal} {Chinese Physics Letters}\ }\textbf {\bibinfo {volume}
  {39}},\ \bibinfo {eid} {100701} (\bibinfo {year} {2022})}\BibitemShut
  {NoStop}%
\bibitem [{\citenamefont {Liu}\ \emph {et~al.}(2023)\citenamefont {Liu},
  \citenamefont {Li}, \citenamefont {Zhang},\ and\ \citenamefont
  {Zhang}}]{PhysRevE.107.L012103}%
  \BibitemOpen
  \bibfield  {author} {\bibinfo {author} {\bibfnamefont {J.}~\bibnamefont
  {Liu}}, \bibinfo {author} {\bibfnamefont {S.}~\bibnamefont {Li}}, \bibinfo
  {author} {\bibfnamefont {J.}~\bibnamefont {Zhang}},\ and\ \bibinfo {author}
  {\bibfnamefont {P.}~\bibnamefont {Zhang}},\ }\bibfield  {title} {\bibinfo
  {title} {Tensor networks for unsupervised machine learning},\ }\href
  {https://doi.org/10.1103/PhysRevE.107.L012103} {\bibfield  {journal}
  {\bibinfo  {journal} {Phys. Rev. E}\ }\textbf {\bibinfo {volume} {107}},\
  \bibinfo {pages} {L012103} (\bibinfo {year} {2023})}\BibitemShut {NoStop}%
\bibitem [{\citenamefont {Meng}\ \emph {et~al.}(2023)\citenamefont {Meng},
  \citenamefont {Zhang}, \citenamefont {Zhang}, \citenamefont {Gao},\ and\
  \citenamefont {Ran}}]{MZZGR23ResMPS}%
  \BibitemOpen
  \bibfield  {author} {\bibinfo {author} {\bibfnamefont {Y.-M.}\ \bibnamefont
  {Meng}}, \bibinfo {author} {\bibfnamefont {J.}~\bibnamefont {Zhang}},
  \bibinfo {author} {\bibfnamefont {P.}~\bibnamefont {Zhang}}, \bibinfo
  {author} {\bibfnamefont {C.}~\bibnamefont {Gao}},\ and\ \bibinfo {author}
  {\bibfnamefont {S.-J.}\ \bibnamefont {Ran}},\ }\bibfield  {title} {\bibinfo
  {title} {{Residual matrix product state for machine learning}},\ }\href
  {https://doi.org/10.21468/SciPostPhys.14.6.142} {\bibfield  {journal}
  {\bibinfo  {journal} {SciPost Phys.}\ }\textbf {\bibinfo {volume} {14}},\
  \bibinfo {pages} {142} (\bibinfo {year} {2023})}\BibitemShut {NoStop}%
\bibitem [{\citenamefont {Chen}\ and\ \citenamefont
  {Barthel}(2024)}]{10517663}%
  \BibitemOpen
  \bibfield  {author} {\bibinfo {author} {\bibfnamefont {H.}~\bibnamefont
  {Chen}}\ and\ \bibinfo {author} {\bibfnamefont {T.}~\bibnamefont {Barthel}},\
  }\bibfield  {title} {\bibinfo {title} {Machine learning with tree tensor
  networks, cp rank constraints, and tensor dropout},\ }\href
  {https://doi.org/10.1109/TPAMI.2024.3396386} {\bibfield  {journal} {\bibinfo
  {journal} {IEEE Transactions on Pattern Analysis and Machine Intelligence}\
  ,\ \bibinfo {pages} {1}} (\bibinfo {year} {2024})}\BibitemShut {NoStop}%
\bibitem [{\citenamefont {Pozas-Kerstjens}\ \emph {et~al.}(2024)\citenamefont
  {Pozas-Kerstjens}, \citenamefont {Hern{\'{a}}ndez-Santana}, \citenamefont
  {Pareja~Monturiol}, \citenamefont {Castrill{\'{o}}n~L{\'{o}}pez},
  \citenamefont {Scarpa}, \citenamefont {Gonz{\'{a}}lez-Guill{\'{e}}n},\ and\
  \citenamefont
  {P{\'{e}}rez-Garc{\'{i}}a}}]{PozasKerstjens2024privacypreserving}%
  \BibitemOpen
  \bibfield  {author} {\bibinfo {author} {\bibfnamefont {A.}~\bibnamefont
  {Pozas-Kerstjens}}, \bibinfo {author} {\bibfnamefont {S.}~\bibnamefont
  {Hern{\'{a}}ndez-Santana}}, \bibinfo {author} {\bibfnamefont {J.~R.}\
  \bibnamefont {Pareja~Monturiol}}, \bibinfo {author} {\bibfnamefont
  {M.}~\bibnamefont {Castrill{\'{o}}n~L{\'{o}}pez}}, \bibinfo {author}
  {\bibfnamefont {G.}~\bibnamefont {Scarpa}}, \bibinfo {author} {\bibfnamefont
  {C.~E.}\ \bibnamefont {Gonz{\'{a}}lez-Guill{\'{e}}n}},\ and\ \bibinfo
  {author} {\bibfnamefont {D.}~\bibnamefont {P{\'{e}}rez-Garc{\'{i}}a}},\
  }\bibfield  {title} {\bibinfo {title} {Privacy-preserving machine learning
  with tensor networks},\ }\href {https://doi.org/10.22331/q-2024-07-25-1425}
  {\bibfield  {journal} {\bibinfo  {journal} {{Quantum}}\ }\textbf {\bibinfo
  {volume} {8}},\ \bibinfo {pages} {1425} (\bibinfo {year} {2024})}\BibitemShut
  {NoStop}%
\bibitem [{\citenamefont {Ran}\ and\ \citenamefont
  {Su}(2023)}]{doi:10.34133/icomputing.0061}%
  \BibitemOpen
  \bibfield  {author} {\bibinfo {author} {\bibfnamefont {S.-J.}\ \bibnamefont
  {Ran}}\ and\ \bibinfo {author} {\bibfnamefont {G.}~\bibnamefont {Su}},\
  }\bibfield  {title} {\bibinfo {title} {Tensor networks for interpretable and
  efficient quantum-inspired machine learning},\ }\href
  {https://doi.org/10.34133/icomputing.0061} {\bibfield  {journal} {\bibinfo
  {journal} {Intelligent Computing}\ }\textbf {\bibinfo {volume} {2}},\
  \bibinfo {pages} {0061} (\bibinfo {year} {2023})}\BibitemShut {NoStop}%
\bibitem [{\citenamefont {F.~Verstraete}\ and\ \citenamefont
  {Cirac}(2008)}]{doi:10.1080/14789940801912366}%
  \BibitemOpen
  \bibfield  {author} {\bibinfo {author} {\bibfnamefont {V.~M.}\ \bibnamefont
  {F.~Verstraete}}\ and\ \bibinfo {author} {\bibfnamefont {J.}~\bibnamefont
  {Cirac}},\ }\bibfield  {title} {\bibinfo {title} {Matrix product states,
  projected entangled pair states, and variational renormalization group
  methods for quantum spin systems},\ }\href
  {https://doi.org/10.1080/14789940801912366} {\bibfield  {journal} {\bibinfo
  {journal} {Advances in Physics}\ }\textbf {\bibinfo {volume} {57}},\ \bibinfo
  {pages} {143} (\bibinfo {year} {2008})}\BibitemShut {NoStop}%
\bibitem [{\citenamefont {Cirac}\ and\ \citenamefont
  {Verstraete}(2009)}]{Cirac_2009}%
  \BibitemOpen
  \bibfield  {author} {\bibinfo {author} {\bibfnamefont {J.~I.}\ \bibnamefont
  {Cirac}}\ and\ \bibinfo {author} {\bibfnamefont {F.}~\bibnamefont
  {Verstraete}},\ }\bibfield  {title} {\bibinfo {title} {Renormalization and
  tensor product states in spin chains and lattices},\ }\href
  {https://doi.org/10.1088/1751-8113/42/50/504004} {\bibfield  {journal}
  {\bibinfo  {journal} {Journal of Physics A: Mathematical and Theoretical}\
  }\textbf {\bibinfo {volume} {42}},\ \bibinfo {pages} {504004} (\bibinfo
  {year} {2009})}\BibitemShut {NoStop}%
\bibitem [{\citenamefont {Orús}(2014)}]{ORUS2014117}%
  \BibitemOpen
  \bibfield  {author} {\bibinfo {author} {\bibfnamefont {R.}~\bibnamefont
  {Orús}},\ }\bibfield  {title} {\bibinfo {title} {A practical introduction to
  tensor networks: Matrix product states and projected entangled pair states},\
  }\href {https://doi.org/https://doi.org/10.1016/j.aop.2014.06.013} {\bibfield
   {journal} {\bibinfo  {journal} {Annals of Physics}\ }\textbf {\bibinfo
  {volume} {349}},\ \bibinfo {pages} {117} (\bibinfo {year}
  {2014})}\BibitemShut {NoStop}%
\bibitem [{\citenamefont {Or{\'u}s}(2019)}]{O19TNrev}%
  \BibitemOpen
  \bibfield  {author} {\bibinfo {author} {\bibfnamefont {R.}~\bibnamefont
  {Or{\'u}s}},\ }\bibfield  {title} {\bibinfo {title} {Tensor networks for
  complex quantum systems},\ }\href {https://doi.org/10.1038/s42254-019-0086-7}
  {\bibfield  {journal} {\bibinfo  {journal} {Nature Reviews Physics}\ }\textbf
  {\bibinfo {volume} {1}},\ \bibinfo {pages} {538} (\bibinfo {year}
  {2019})}\BibitemShut {NoStop}%
\bibitem [{\citenamefont {Ran}\ \emph {et~al.}(2020)\citenamefont {Ran},
  \citenamefont {Tirrito}, \citenamefont {Peng}, \citenamefont {Chen},
  \citenamefont {Tagliacozzo}, \citenamefont {Su},\ and\ \citenamefont
  {Lewenstein}}]{RTPC+17TNrev}%
  \BibitemOpen
  \bibfield  {author} {\bibinfo {author} {\bibfnamefont {S.-J.}\ \bibnamefont
  {Ran}}, \bibinfo {author} {\bibfnamefont {E.}~\bibnamefont {Tirrito}},
  \bibinfo {author} {\bibfnamefont {C.}~\bibnamefont {Peng}}, \bibinfo {author}
  {\bibfnamefont {X.}~\bibnamefont {Chen}}, \bibinfo {author} {\bibfnamefont
  {L.}~\bibnamefont {Tagliacozzo}}, \bibinfo {author} {\bibfnamefont
  {G.}~\bibnamefont {Su}},\ and\ \bibinfo {author} {\bibfnamefont
  {M.}~\bibnamefont {Lewenstein}},\ }\href
  {https://doi.org/10.1007/978-3-030-34489-4} {\emph {\bibinfo {title} {Tensor
  Network Contractions: Methods and Applications to Quantum Many-Body
  Systems}}}\ (\bibinfo  {publisher} {Springer, Cham},\ \bibinfo {year}
  {2020})\BibitemShut {NoStop}%
\bibitem [{\citenamefont {Cirac}\ \emph {et~al.}(2021)\citenamefont {Cirac},
  \citenamefont {P\'erez-Garc\'{\i}a}, \citenamefont {Schuch},\ and\
  \citenamefont {Verstraete}}]{cirac2021matrix}%
  \BibitemOpen
  \bibfield  {author} {\bibinfo {author} {\bibfnamefont {J.~I.}\ \bibnamefont
  {Cirac}}, \bibinfo {author} {\bibfnamefont {D.}~\bibnamefont
  {P\'erez-Garc\'{\i}a}}, \bibinfo {author} {\bibfnamefont {N.}~\bibnamefont
  {Schuch}},\ and\ \bibinfo {author} {\bibfnamefont {F.}~\bibnamefont
  {Verstraete}},\ }\bibfield  {title} {\bibinfo {title} {Matrix product states
  and projected entangled pair states: Concepts, symmetries, theorems},\ }\href
  {https://doi.org/10.1103/RevModPhys.93.045003} {\bibfield  {journal}
  {\bibinfo  {journal} {Rev. Mod. Phys.}\ }\textbf {\bibinfo {volume} {93}},\
  \bibinfo {pages} {045003} (\bibinfo {year} {2021})}\BibitemShut {NoStop}%
\bibitem [{\citenamefont {Vidal}\ \emph {et~al.}(2003)\citenamefont {Vidal},
  \citenamefont {Latorre}, \citenamefont {Rico},\ and\ \citenamefont
  {Kitaev}}]{PhysRevLett.90.227902}%
  \BibitemOpen
  \bibfield  {author} {\bibinfo {author} {\bibfnamefont {G.}~\bibnamefont
  {Vidal}}, \bibinfo {author} {\bibfnamefont {J.~I.}\ \bibnamefont {Latorre}},
  \bibinfo {author} {\bibfnamefont {E.}~\bibnamefont {Rico}},\ and\ \bibinfo
  {author} {\bibfnamefont {A.}~\bibnamefont {Kitaev}},\ }\bibfield  {title}
  {\bibinfo {title} {Entanglement in quantum critical phenomena},\ }\href
  {https://doi.org/10.1103/PhysRevLett.90.227902} {\bibfield  {journal}
  {\bibinfo  {journal} {Phys. Rev. Lett.}\ }\textbf {\bibinfo {volume} {90}},\
  \bibinfo {pages} {227902} (\bibinfo {year} {2003})}\BibitemShut {NoStop}%
\bibitem [{\citenamefont {Pollmann}\ \emph {et~al.}(2009)\citenamefont
  {Pollmann}, \citenamefont {Mukerjee}, \citenamefont {Turner},\ and\
  \citenamefont {Moore}}]{PhysRevLett.102.255701}%
  \BibitemOpen
  \bibfield  {author} {\bibinfo {author} {\bibfnamefont {F.}~\bibnamefont
  {Pollmann}}, \bibinfo {author} {\bibfnamefont {S.}~\bibnamefont {Mukerjee}},
  \bibinfo {author} {\bibfnamefont {A.~M.}\ \bibnamefont {Turner}},\ and\
  \bibinfo {author} {\bibfnamefont {J.~E.}\ \bibnamefont {Moore}},\ }\bibfield
  {title} {\bibinfo {title} {Theory of finite-entanglement scaling at
  one-dimensional quantum critical points},\ }\href
  {https://doi.org/10.1103/PhysRevLett.102.255701} {\bibfield  {journal}
  {\bibinfo  {journal} {Phys. Rev. Lett.}\ }\textbf {\bibinfo {volume} {102}},\
  \bibinfo {pages} {255701} (\bibinfo {year} {2009})}\BibitemShut {NoStop}%
\bibitem [{\citenamefont {Schuch}\ \emph {et~al.}(2008)\citenamefont {Schuch},
  \citenamefont {Wolf}, \citenamefont {Verstraete},\ and\ \citenamefont
  {Cirac}}]{SWVC08MPSent}%
  \BibitemOpen
  \bibfield  {author} {\bibinfo {author} {\bibfnamefont {N.}~\bibnamefont
  {Schuch}}, \bibinfo {author} {\bibfnamefont {M.~M.}\ \bibnamefont {Wolf}},
  \bibinfo {author} {\bibfnamefont {F.}~\bibnamefont {Verstraete}},\ and\
  \bibinfo {author} {\bibfnamefont {J.~I.}\ \bibnamefont {Cirac}},\ }\bibfield
  {title} {\bibinfo {title} {Entropy scaling and simulability by matrix product
  states},\ }\href {https://doi.org/10.1103/PhysRevLett.100.030504} {\bibfield
  {journal} {\bibinfo  {journal} {Phys. Rev. Lett.}\ }\textbf {\bibinfo
  {volume} {100}},\ \bibinfo {pages} {030504} (\bibinfo {year}
  {2008})}\BibitemShut {NoStop}%
\bibitem [{\citenamefont {Tagliacozzo}\ \emph {et~al.}(2008)\citenamefont
  {Tagliacozzo}, \citenamefont {de~Oliveira}, \citenamefont {Iblisdir},\ and\
  \citenamefont {Latorre}}]{TOIL08EntScaling}%
  \BibitemOpen
  \bibfield  {author} {\bibinfo {author} {\bibfnamefont {L.}~\bibnamefont
  {Tagliacozzo}}, \bibinfo {author} {\bibfnamefont {T.~R.}\ \bibnamefont
  {de~Oliveira}}, \bibinfo {author} {\bibfnamefont {S.}~\bibnamefont
  {Iblisdir}},\ and\ \bibinfo {author} {\bibfnamefont {J.~I.}\ \bibnamefont
  {Latorre}},\ }\bibfield  {title} {\bibinfo {title} {{Scaling of entanglement
  support for matrix product states}},\ }\href
  {https://doi.org/10.1103/PhysRevB.78.024410} {\bibfield  {journal} {\bibinfo
  {journal} {Phys. Rev. B}\ }\textbf {\bibinfo {volume} {78}},\ \bibinfo
  {pages} {024410} (\bibinfo {year} {2008})}\BibitemShut {NoStop}%
\bibitem [{\citenamefont {P\'{e}rez-Garc\'ia}\ \emph
  {et~al.}(2007)\citenamefont {P\'{e}rez-Garc\'ia}, \citenamefont {Verstraete},
  \citenamefont {Wolf},\ and\ \citenamefont {Cirac}}]{PVWC07MPSRev}%
  \BibitemOpen
  \bibfield  {author} {\bibinfo {author} {\bibfnamefont {D.}~\bibnamefont
  {P\'{e}rez-Garc\'ia}}, \bibinfo {author} {\bibfnamefont {F.}~\bibnamefont
  {Verstraete}}, \bibinfo {author} {\bibfnamefont {M.~M.}\ \bibnamefont
  {Wolf}},\ and\ \bibinfo {author} {\bibfnamefont {J.~I.}\ \bibnamefont
  {Cirac}},\ }\bibfield  {title} {\bibinfo {title} {{Matrix Product State
  Representations}},\ }\href@noop {} {\bibfield  {journal} {\bibinfo  {journal}
  {Quantum Inf. Comput.}\ }\textbf {\bibinfo {volume} {7}},\ \bibinfo {pages}
  {401} (\bibinfo {year} {2007})}\BibitemShut {NoStop}%
\bibitem [{\citenamefont {Hestness}\ \emph {et~al.}(2017)\citenamefont
  {Hestness}, \citenamefont {Narang}, \citenamefont {Ardalani}, \citenamefont
  {Diamos}, \citenamefont {Jun}, \citenamefont {Kianinejad}, \citenamefont
  {Patwary}, \citenamefont {Yang},\ and\ \citenamefont
  {Zhou}}]{hestness2017deeplearningscalingpredictable}%
  \BibitemOpen
  \bibfield  {author} {\bibinfo {author} {\bibfnamefont {J.}~\bibnamefont
  {Hestness}}, \bibinfo {author} {\bibfnamefont {S.}~\bibnamefont {Narang}},
  \bibinfo {author} {\bibfnamefont {N.}~\bibnamefont {Ardalani}}, \bibinfo
  {author} {\bibfnamefont {G.}~\bibnamefont {Diamos}}, \bibinfo {author}
  {\bibfnamefont {H.}~\bibnamefont {Jun}}, \bibinfo {author} {\bibfnamefont
  {H.}~\bibnamefont {Kianinejad}}, \bibinfo {author} {\bibfnamefont {M.~M.~A.}\
  \bibnamefont {Patwary}}, \bibinfo {author} {\bibfnamefont {Y.}~\bibnamefont
  {Yang}},\ and\ \bibinfo {author} {\bibfnamefont {Y.}~\bibnamefont {Zhou}},\
  }\href {https://arxiv.org/abs/1712.00409} {\bibinfo {title} {Deep learning
  scaling is predictable, empirically}} (\bibinfo {year} {2017}),\ \Eprint
  {https://arxiv.org/abs/1712.00409} {arXiv:1712.00409 [cs.LG]} \BibitemShut
  {NoStop}%
\bibitem [{\citenamefont {Kaplan}\ \emph {et~al.}(2020)\citenamefont {Kaplan},
  \citenamefont {McCandlish}, \citenamefont {Henighan}, \citenamefont {Brown},
  \citenamefont {Chess}, \citenamefont {Child}, \citenamefont {Gray},
  \citenamefont {Radford}, \citenamefont {Wu},\ and\ \citenamefont
  {Amodei}}]{kaplan2020scalinglawsneurallanguage}%
  \BibitemOpen
  \bibfield  {author} {\bibinfo {author} {\bibfnamefont {J.}~\bibnamefont
  {Kaplan}}, \bibinfo {author} {\bibfnamefont {S.}~\bibnamefont {McCandlish}},
  \bibinfo {author} {\bibfnamefont {T.}~\bibnamefont {Henighan}}, \bibinfo
  {author} {\bibfnamefont {T.~B.}\ \bibnamefont {Brown}}, \bibinfo {author}
  {\bibfnamefont {B.}~\bibnamefont {Chess}}, \bibinfo {author} {\bibfnamefont
  {R.}~\bibnamefont {Child}}, \bibinfo {author} {\bibfnamefont
  {S.}~\bibnamefont {Gray}}, \bibinfo {author} {\bibfnamefont {A.}~\bibnamefont
  {Radford}}, \bibinfo {author} {\bibfnamefont {J.}~\bibnamefont {Wu}},\ and\
  \bibinfo {author} {\bibfnamefont {D.}~\bibnamefont {Amodei}},\ }\href
  {https://arxiv.org/abs/2001.08361} {\bibinfo {title} {Scaling laws for neural
  language models}} (\bibinfo {year} {2020}),\ \Eprint
  {https://arxiv.org/abs/2001.08361} {arXiv:2001.08361 [cs.LG]} \BibitemShut
  {NoStop}%
\bibitem [{\citenamefont {Henighan}\ \emph {et~al.}(2020)\citenamefont
  {Henighan}, \citenamefont {Kaplan}, \citenamefont {Katz}, \citenamefont
  {Chen}, \citenamefont {Hesse}, \citenamefont {Jackson}, \citenamefont {Jun},
  \citenamefont {Brown}, \citenamefont {Dhariwal}, \citenamefont {Gray},
  \citenamefont {Hallacy}, \citenamefont {Mann}, \citenamefont {Radford},
  \citenamefont {Ramesh}, \citenamefont {Ryder}, \citenamefont {Ziegler},
  \citenamefont {Schulman}, \citenamefont {Amodei},\ and\ \citenamefont
  {McCandlish}}]{henighan2020scalinglawsautoregressivegenerative}%
  \BibitemOpen
  \bibfield  {author} {\bibinfo {author} {\bibfnamefont {T.}~\bibnamefont
  {Henighan}}, \bibinfo {author} {\bibfnamefont {J.}~\bibnamefont {Kaplan}},
  \bibinfo {author} {\bibfnamefont {M.}~\bibnamefont {Katz}}, \bibinfo {author}
  {\bibfnamefont {M.}~\bibnamefont {Chen}}, \bibinfo {author} {\bibfnamefont
  {C.}~\bibnamefont {Hesse}}, \bibinfo {author} {\bibfnamefont
  {J.}~\bibnamefont {Jackson}}, \bibinfo {author} {\bibfnamefont
  {H.}~\bibnamefont {Jun}}, \bibinfo {author} {\bibfnamefont {T.~B.}\
  \bibnamefont {Brown}}, \bibinfo {author} {\bibfnamefont {P.}~\bibnamefont
  {Dhariwal}}, \bibinfo {author} {\bibfnamefont {S.}~\bibnamefont {Gray}},
  \bibinfo {author} {\bibfnamefont {C.}~\bibnamefont {Hallacy}}, \bibinfo
  {author} {\bibfnamefont {B.}~\bibnamefont {Mann}}, \bibinfo {author}
  {\bibfnamefont {A.}~\bibnamefont {Radford}}, \bibinfo {author} {\bibfnamefont
  {A.}~\bibnamefont {Ramesh}}, \bibinfo {author} {\bibfnamefont
  {N.}~\bibnamefont {Ryder}}, \bibinfo {author} {\bibfnamefont {D.~M.}\
  \bibnamefont {Ziegler}}, \bibinfo {author} {\bibfnamefont {J.}~\bibnamefont
  {Schulman}}, \bibinfo {author} {\bibfnamefont {D.}~\bibnamefont {Amodei}},\
  and\ \bibinfo {author} {\bibfnamefont {S.}~\bibnamefont {McCandlish}},\
  }\href {https://arxiv.org/abs/2010.14701} {\bibinfo {title} {Scaling laws for
  autoregressive generative modeling}} (\bibinfo {year} {2020}),\ \Eprint
  {https://arxiv.org/abs/2010.14701} {arXiv:2010.14701 [cs.LG]} \BibitemShut
  {NoStop}%
\bibitem [{\citenamefont {Zhai}\ \emph {et~al.}(2022)\citenamefont {Zhai},
  \citenamefont {Kolesnikov}, \citenamefont {Houlsby},\ and\ \citenamefont
  {Beyer}}]{Zhai_2022_CVPR}%
  \BibitemOpen
  \bibfield  {author} {\bibinfo {author} {\bibfnamefont {X.}~\bibnamefont
  {Zhai}}, \bibinfo {author} {\bibfnamefont {A.}~\bibnamefont {Kolesnikov}},
  \bibinfo {author} {\bibfnamefont {N.}~\bibnamefont {Houlsby}},\ and\ \bibinfo
  {author} {\bibfnamefont {L.}~\bibnamefont {Beyer}},\ }\bibfield  {title}
  {\bibinfo {title} {Scaling vision transformers},\ }in\ \href@noop {} {\emph
  {\bibinfo {booktitle} {Proceedings of the IEEE/CVF Conference on Computer
  Vision and Pattern Recognition (CVPR)}}}\ (\bibinfo {year} {2022})\ pp.\
  \bibinfo {pages} {12104--12113}\BibitemShut {NoStop}%
\bibitem [{\citenamefont {Hoffmann}\ \emph {et~al.}(2022)\citenamefont
  {Hoffmann}, \citenamefont {Borgeaud}, \citenamefont {Mensch}, \citenamefont
  {Buchatskaya}, \citenamefont {Cai}, \citenamefont {Rutherford}, \citenamefont
  {de~Las~Casas}, \citenamefont {Hendricks}, \citenamefont {Welbl},
  \citenamefont {Clark}, \citenamefont {Hennigan}, \citenamefont {Noland},
  \citenamefont {Millican}, \citenamefont {van~den Driessche}, \citenamefont
  {Damoc}, \citenamefont {Guy}, \citenamefont {Osindero}, \citenamefont
  {Simonyan}, \citenamefont {Elsen}, \citenamefont {Rae}, \citenamefont
  {Vinyals},\ and\ \citenamefont
  {Sifre}}]{hoffmann2022trainingcomputeoptimallargelanguage}%
  \BibitemOpen
  \bibfield  {author} {\bibinfo {author} {\bibfnamefont {J.}~\bibnamefont
  {Hoffmann}}, \bibinfo {author} {\bibfnamefont {S.}~\bibnamefont {Borgeaud}},
  \bibinfo {author} {\bibfnamefont {A.}~\bibnamefont {Mensch}}, \bibinfo
  {author} {\bibfnamefont {E.}~\bibnamefont {Buchatskaya}}, \bibinfo {author}
  {\bibfnamefont {T.}~\bibnamefont {Cai}}, \bibinfo {author} {\bibfnamefont
  {E.}~\bibnamefont {Rutherford}}, \bibinfo {author} {\bibfnamefont
  {D.}~\bibnamefont {de~Las~Casas}}, \bibinfo {author} {\bibfnamefont {L.~A.}\
  \bibnamefont {Hendricks}}, \bibinfo {author} {\bibfnamefont {J.}~\bibnamefont
  {Welbl}}, \bibinfo {author} {\bibfnamefont {A.}~\bibnamefont {Clark}},
  \bibinfo {author} {\bibfnamefont {T.}~\bibnamefont {Hennigan}}, \bibinfo
  {author} {\bibfnamefont {E.}~\bibnamefont {Noland}}, \bibinfo {author}
  {\bibfnamefont {K.}~\bibnamefont {Millican}}, \bibinfo {author}
  {\bibfnamefont {G.}~\bibnamefont {van~den Driessche}}, \bibinfo {author}
  {\bibfnamefont {B.}~\bibnamefont {Damoc}}, \bibinfo {author} {\bibfnamefont
  {A.}~\bibnamefont {Guy}}, \bibinfo {author} {\bibfnamefont {S.}~\bibnamefont
  {Osindero}}, \bibinfo {author} {\bibfnamefont {K.}~\bibnamefont {Simonyan}},
  \bibinfo {author} {\bibfnamefont {E.}~\bibnamefont {Elsen}}, \bibinfo
  {author} {\bibfnamefont {J.~W.}\ \bibnamefont {Rae}}, \bibinfo {author}
  {\bibfnamefont {O.}~\bibnamefont {Vinyals}},\ and\ \bibinfo {author}
  {\bibfnamefont {L.}~\bibnamefont {Sifre}},\ }\href
  {https://arxiv.org/abs/2203.15556} {\bibinfo {title} {Training
  compute-optimal large language models}} (\bibinfo {year} {2022}),\ \Eprint
  {https://arxiv.org/abs/2203.15556} {arXiv:2203.15556 [cs.CL]} \BibitemShut
  {NoStop}%
\bibitem [{\citenamefont {Sorscher}\ \emph {et~al.}(2022)\citenamefont
  {Sorscher}, \citenamefont {Geirhos}, \citenamefont {Shekhar}, \citenamefont
  {Ganguli},\ and\ \citenamefont {Morcos}}]{NEURIPS2022_7b75da9b}%
  \BibitemOpen
  \bibfield  {author} {\bibinfo {author} {\bibfnamefont {B.}~\bibnamefont
  {Sorscher}}, \bibinfo {author} {\bibfnamefont {R.}~\bibnamefont {Geirhos}},
  \bibinfo {author} {\bibfnamefont {S.}~\bibnamefont {Shekhar}}, \bibinfo
  {author} {\bibfnamefont {S.}~\bibnamefont {Ganguli}},\ and\ \bibinfo {author}
  {\bibfnamefont {A.}~\bibnamefont {Morcos}},\ }\bibfield  {title} {\bibinfo
  {title} {Beyond neural scaling laws: beating power law scaling via data
  pruning},\ }in\ \href
  {https://proceedings.neurips.cc/paper_files/paper/2022/file/7b75da9b61eda40fa35453ee5d077df6-Paper-Conference.pdf}
  {\emph {\bibinfo {booktitle} {Advances in Neural Information Processing
  Systems}}},\ Vol.~\bibinfo {volume} {35},\ \bibinfo {editor} {edited by\
  \bibinfo {editor} {\bibfnamefont {S.}~\bibnamefont {Koyejo}}, \bibinfo
  {editor} {\bibfnamefont {S.}~\bibnamefont {Mohamed}}, \bibinfo {editor}
  {\bibfnamefont {A.}~\bibnamefont {Agarwal}}, \bibinfo {editor} {\bibfnamefont
  {D.}~\bibnamefont {Belgrave}}, \bibinfo {editor} {\bibfnamefont
  {K.}~\bibnamefont {Cho}},\ and\ \bibinfo {editor} {\bibfnamefont
  {A.}~\bibnamefont {Oh}}}\ (\bibinfo  {publisher} {Curran Associates, Inc.},\
  \bibinfo {year} {2022})\ pp.\ \bibinfo {pages} {19523--19536}\BibitemShut
  {NoStop}%
\bibitem [{\citenamefont {Muennighoff}\ \emph {et~al.}(2023)\citenamefont
  {Muennighoff}, \citenamefont {Rush}, \citenamefont {Barak}, \citenamefont
  {Le~Scao}, \citenamefont {Tazi}, \citenamefont {Piktus}, \citenamefont
  {Pyysalo}, \citenamefont {Wolf},\ and\ \citenamefont
  {Raffel}}]{NEURIPS2023_9d89448b}%
  \BibitemOpen
  \bibfield  {author} {\bibinfo {author} {\bibfnamefont {N.}~\bibnamefont
  {Muennighoff}}, \bibinfo {author} {\bibfnamefont {A.}~\bibnamefont {Rush}},
  \bibinfo {author} {\bibfnamefont {B.}~\bibnamefont {Barak}}, \bibinfo
  {author} {\bibfnamefont {T.}~\bibnamefont {Le~Scao}}, \bibinfo {author}
  {\bibfnamefont {N.}~\bibnamefont {Tazi}}, \bibinfo {author} {\bibfnamefont
  {A.}~\bibnamefont {Piktus}}, \bibinfo {author} {\bibfnamefont
  {S.}~\bibnamefont {Pyysalo}}, \bibinfo {author} {\bibfnamefont
  {T.}~\bibnamefont {Wolf}},\ and\ \bibinfo {author} {\bibfnamefont {C.~A.}\
  \bibnamefont {Raffel}},\ }\bibfield  {title} {\bibinfo {title} {Scaling
  data-constrained language models},\ }in\ \href
  {https://proceedings.neurips.cc/paper_files/paper/2023/file/9d89448b63ce1e2e8dc7af72c984c196-Paper-Conference.pdf}
  {\emph {\bibinfo {booktitle} {Advances in Neural Information Processing
  Systems}}},\ Vol.~\bibinfo {volume} {36},\ \bibinfo {editor} {edited by\
  \bibinfo {editor} {\bibfnamefont {A.}~\bibnamefont {Oh}}, \bibinfo {editor}
  {\bibfnamefont {T.}~\bibnamefont {Naumann}}, \bibinfo {editor} {\bibfnamefont
  {A.}~\bibnamefont {Globerson}}, \bibinfo {editor} {\bibfnamefont
  {K.}~\bibnamefont {Saenko}}, \bibinfo {editor} {\bibfnamefont
  {M.}~\bibnamefont {Hardt}},\ and\ \bibinfo {editor} {\bibfnamefont
  {S.}~\bibnamefont {Levine}}}\ (\bibinfo  {publisher} {Curran Associates,
  Inc.},\ \bibinfo {year} {2023})\ pp.\ \bibinfo {pages}
  {50358--50376}\BibitemShut {NoStop}%
\bibitem [{\citenamefont {Gao}\ \emph {et~al.}(2023)\citenamefont {Gao},
  \citenamefont {Schulman},\ and\ \citenamefont {Hilton}}]{pmlr-v202-gao23h}%
  \BibitemOpen
  \bibfield  {author} {\bibinfo {author} {\bibfnamefont {L.}~\bibnamefont
  {Gao}}, \bibinfo {author} {\bibfnamefont {J.}~\bibnamefont {Schulman}},\ and\
  \bibinfo {author} {\bibfnamefont {J.}~\bibnamefont {Hilton}},\ }\bibfield
  {title} {\bibinfo {title} {Scaling laws for reward model overoptimization},\
  }in\ \href {https://proceedings.mlr.press/v202/gao23h.html} {\emph {\bibinfo
  {booktitle} {Proceedings of the 40th International Conference on Machine
  Learning}}},\ \bibinfo {series} {Proceedings of Machine Learning Research},
  Vol.\ \bibinfo {volume} {202},\ \bibinfo {editor} {edited by\ \bibinfo
  {editor} {\bibfnamefont {A.}~\bibnamefont {Krause}}, \bibinfo {editor}
  {\bibfnamefont {E.}~\bibnamefont {Brunskill}}, \bibinfo {editor}
  {\bibfnamefont {K.}~\bibnamefont {Cho}}, \bibinfo {editor} {\bibfnamefont
  {B.}~\bibnamefont {Engelhardt}}, \bibinfo {editor} {\bibfnamefont
  {S.}~\bibnamefont {Sabato}},\ and\ \bibinfo {editor} {\bibfnamefont
  {J.}~\bibnamefont {Scarlett}}}\ (\bibinfo  {publisher} {PMLR},\ \bibinfo
  {year} {2023})\ pp.\ \bibinfo {pages} {10835--10866}\BibitemShut {NoStop}%
\bibitem [{\citenamefont {Convy}\ \emph {et~al.}(2022)\citenamefont {Convy},
  \citenamefont {Huggins}, \citenamefont {Liao},\ and\ \citenamefont
  {Whaley}}]{Convy_2022}%
  \BibitemOpen
  \bibfield  {author} {\bibinfo {author} {\bibfnamefont {I.}~\bibnamefont
  {Convy}}, \bibinfo {author} {\bibfnamefont {W.}~\bibnamefont {Huggins}},
  \bibinfo {author} {\bibfnamefont {H.}~\bibnamefont {Liao}},\ and\ \bibinfo
  {author} {\bibfnamefont {K.~B.}\ \bibnamefont {Whaley}},\ }\bibfield  {title}
  {\bibinfo {title} {Mutual information scaling for tensor network machine
  learning},\ }\href {https://doi.org/10.1088/2632-2153/ac44a9} {\bibfield
  {journal} {\bibinfo  {journal} {Machine Learning: Science and Technology}\
  }\textbf {\bibinfo {volume} {3}},\ \bibinfo {pages} {015017} (\bibinfo {year}
  {2022})}\BibitemShut {NoStop}%
\bibitem [{\citenamefont {Xiao}\ \emph {et~al.}(2017)\citenamefont {Xiao},
  \citenamefont {Rasul},\ and\ \citenamefont {Vollgraf}}]{fMNIST}%
  \BibitemOpen
  \bibfield  {author} {\bibinfo {author} {\bibfnamefont {H.}~\bibnamefont
  {Xiao}}, \bibinfo {author} {\bibfnamefont {K.}~\bibnamefont {Rasul}},\ and\
  \bibinfo {author} {\bibfnamefont {R.}~\bibnamefont {Vollgraf}},\ }\bibfield
  {title} {\bibinfo {title} {Fashion-mnist: a novel image dataset for
  benchmarking machine learning algorithms},\ }\href@noop {} {\  (\bibinfo
  {year} {2017})},\ \Eprint {https://arxiv.org/abs/arXiv:1708.07747}
  {arXiv:1708.07747} \BibitemShut {NoStop}%
\bibitem [{SM()}]{SM}%
  \BibitemOpen
  \href@noop {} {}\bibinfo {note} {In the appendices, additional results are
  provided to demonstrate the universality of the scaling laws. The scaling
  against the number of training samples is investigated. By designing a
  multi-spin quantum feature map, we further show the influence to the
  representation and generalization powers by tuning
  orthogonality.}\BibitemShut {Stop}%
\bibitem [{\citenamefont {Deng}(2012)}]{MNIST_6296535}%
  \BibitemOpen
  \bibfield  {author} {\bibinfo {author} {\bibfnamefont {L.}~\bibnamefont
  {Deng}},\ }\bibfield  {title} {\bibinfo {title} {The mnist database of
  handwritten digit images for machine learning research [best of the web]},\
  }\href {https://doi.org/10.1109/MSP.2012.2211477} {\bibfield  {journal}
  {\bibinfo  {journal} {IEEE Signal Processing Magazine}\ }\textbf {\bibinfo
  {volume} {29}},\ \bibinfo {pages} {141} (\bibinfo {year} {2012})}\BibitemShut
  {NoStop}%
\bibitem [{\citenamefont {Clanuwat}\ \emph {et~al.}(2018)\citenamefont
  {Clanuwat}, \citenamefont {Bober-Irizar}, \citenamefont {Kitamoto},
  \citenamefont {Lamb}, \citenamefont {Yamamoto},\ and\ \citenamefont
  {Ha}}]{clanuwat2018deep}%
  \BibitemOpen
  \bibfield  {author} {\bibinfo {author} {\bibfnamefont {T.}~\bibnamefont
  {Clanuwat}}, \bibinfo {author} {\bibfnamefont {M.}~\bibnamefont
  {Bober-Irizar}}, \bibinfo {author} {\bibfnamefont {A.}~\bibnamefont
  {Kitamoto}}, \bibinfo {author} {\bibfnamefont {A.}~\bibnamefont {Lamb}},
  \bibinfo {author} {\bibfnamefont {K.}~\bibnamefont {Yamamoto}},\ and\
  \bibinfo {author} {\bibfnamefont {D.}~\bibnamefont {Ha}},\ }\href@noop {}
  {\bibinfo {title} {Deep learning for classical japanese literature}}
  (\bibinfo {year} {2018}),\ \Eprint {https://arxiv.org/abs/arXiv:1812.01718}
  {arXiv:1812.01718} \BibitemShut {NoStop}%
\bibitem [{\citenamefont {Reyes-Ortiz}\ and\ \citenamefont
  {Parra}(2013)}]{human_activity_recognition_using_smartphones_240}%
  \BibitemOpen
  \bibfield  {author} {\bibinfo {author} {\bibfnamefont {A.~D. G. A. O.~L.}\
  \bibnamefont {Reyes-Ortiz}, \bibfnamefont {Jorge}}\ and\ \bibinfo {author}
  {\bibfnamefont {X.}~\bibnamefont {Parra}},\ }\href@noop {} {\bibinfo {title}
  {{Human Activity Recognition Using Smartphones}}},\ \bibinfo {howpublished}
  {UCI Machine Learning Repository} (\bibinfo {year} {2013}),\ \bibinfo {note}
  {{DOI}: https://doi.org/10.24432/C54S4K}\BibitemShut {NoStop}%
\bibitem [{\citenamefont {May}(1976)}]{may1976simple}%
  \BibitemOpen
  \bibfield  {author} {\bibinfo {author} {\bibfnamefont {R.~M.}\ \bibnamefont
  {May}},\ }\bibfield  {title} {\bibinfo {title} {Simple mathematical models
  with very complicated dynamics},\ }\href {https://doi.org/10.1038/261459a0}
  {\bibfield  {journal} {\bibinfo  {journal} {Nature}\ }\textbf {\bibinfo
  {volume} {261}},\ \bibinfo {pages} {459} (\bibinfo {year}
  {1976})}\BibitemShut {NoStop}%
\bibitem [{dry(2020)}]{dry_bean_602}%
  \BibitemOpen
  \href@noop {} {\bibinfo {title} {{Dry Bean}}},\ \bibinfo {howpublished} {UCI
  Machine Learning Repository} (\bibinfo {year} {2020}),\ \bibinfo {note}
  {{DOI}: https://doi.org/10.24432/C50S4B}\BibitemShut {NoStop}%
\end{thebibliography}

\end{document}